\documentclass[10pt,conference]{IEEEtran}
\usepackage{cite}
\usepackage{soul}
\usepackage{amsmath,amssymb,amsfonts}
\usepackage{algorithmic}
\usepackage{graphicx}
\usepackage{textcomp}
\usepackage{xcolor}
\usepackage{subfigure}
\usepackage{xspace}
\usepackage{mathtools}
\usepackage{enumitem}
\usepackage{multirow}
\usepackage{tabularx}
\usepackage{arydshln}
\usepackage{adjustbox}
\usepackage{tikz}
\usepackage{caption}
\usepackage{verbatim}
\usepackage{mdframed}

\def\BibTeX{{\rm B\kern-.05em{\sc i\kern-.025em b}\kern-.08em
    T\kern-.1667em\lower.7ex\hbox{E}\kern-.125emX}}

\newcommand{\codebert}{CodeBERT\xspace}
\newcommand{\graphcodebert}{GraphCodeBERT\xspace}
\newcommand{\plbart}{PLBART\xspace}
\newcommand{\codetf}{CodeT5\xspace}
\newcommand{\codegen}{CodeGen\xspace}
\newcommand{\codex}{Codex\xspace}

\newcommand{\cure}{CURE\xspace}

\newcommand{\recoder}{Recoder\xspace}

\newcommand{\rewardrepair}{RewardRepair\xspace}

\newcommand{\defects}{Defects4J\xspace}
\newcommand{\defectsold}{Defects4J v1.2\xspace}
\newcommand{\defectsnew}{Defects4J v2.0\xspace}
\newcommand{\quixbugs}{QuixBugs\xspace}

\newcommand{\todoc}[2]{{\textcolor{#1}{\textbf{#2}}}}

\newcommand{\todored}[1]{{\todoc{red}{\textbf{[[#1]]}}}}
\newcommand{\todogreen}[1]{\todoc{green}{\textbf{[[#1]]}}}
\newcommand{\todoblue}[1]{\todoc{blue}{\textbf{[[#1]]}}}

\newcommand{\todobrown}[1]{\todoc{brown}{\textbf{[[#1]]}}}

\newcommand{\todo}[1]{\todored{TODO: #1}}

\newcommand{\lin}[1]{\todoblue{Lin: #1}}
\newcommand{\nan}[1]{\todobrown{Nan: #1}}
\newcommand{\thibaud}[1]{\todogreen{Thibaud: #1}}

\renewcommand{\todoc}[2]{\relax}
\newcommand{\rev}[1]{{#1}}

\newcommand{\code}[1]{\texttt{\small #1}} 

\newcounter{finding}
\newcommand{\finding}[1]{\refstepcounter{finding}
	\begin{mdframed}[linecolor=gray,roundcorner=12pt,backgroundcolor=gray!15,linewidth=3pt,innerleftmargin=2pt, leftmargin=0cm,rightmargin=0cm,topline=false,bottomline=false,rightline=false]
	\textbf{Finding \arabic{finding}:} #1
	\end{mdframed}
}

\newcommand{\distance}{8pt}
\begin{document}

\title{Impact of Code Language Models on Automated Program Repair}

\author{
\IEEEauthorblockN{Nan Jiang}
\IEEEauthorblockA{
\textit{Purdue University}\\
West Lafayette, USA \\
jiang719@purdue.edu}
\and
\IEEEauthorblockN{Kevin Liu}
\IEEEauthorblockA{
\textit{Lynbrook High School}\\
San Jose, USA \\
kevin.bx.liu@gmail.com}
\and
\IEEEauthorblockN{Thibaud Lutellier}
\IEEEauthorblockA{
\textit{University of Alberta}\\
Alberta, Canada \\
lutellie@ualberta.ca}
\and
\IEEEauthorblockN{Lin Tan}
\IEEEauthorblockA{
\textit{Purdue University}\\
West Lafayette, USA \\
lintan@purdue.edu}
}

\maketitle

\begin{abstract}
Automated program repair (APR) aims to help developers improve software reliability by generating patches for buggy programs. Although many code language models (CLM) are developed and effective in many software tasks such as code completion, there has been little comprehensive, in-depth work to evaluate CLMs' fixing capabilities and to fine-tune CLMs for the APR task.

Firstly, this work is the first to evaluate ten CLMs on four APR benchmarks, which shows that surprisingly, the best CLM, as is, fixes 72\% more bugs than the state-of-the-art deep-learning (DL)-based APR techniques. 
Secondly, one of the four APR benchmarks was created by us in this paper to avoid data leaking for a fair evaluation. Thirdly, it is the first work to fine-tune CLMs with APR training data, which shows that fine-tuning brings 31\%--1,267\% improvement to CLMs and enables them to fix 46\%--164\% more bugs than existing DL-based APR techniques. 
Fourthly, this work studies the impact of buggy lines, showing that CLMs, as is, cannot make good use of the buggy lines to fix bugs, yet fine-tuned CLMs could potentially over-rely on buggy lines. Lastly, this work analyzes the size, time, and memory efficiency of different CLMs.

This work shows promising directions for the APR domain, such as fine-tuning CLMs with APR-specific designs, and also raises awareness of fair and comprehensive evaluations of CLMs and calls for more transparent reporting of open-source repositories used in the pre-training data to address the data leaking problem.

\end{abstract}


\begin{tikzpicture}[remember picture,overlay]
    \node[anchor=south,yshift=765pt] at (current page.south) {\parbox{\dimexpr 1\textwidth-\fboxsep-\fboxrule\relax}{\centering 2023 IEEE/ACM 45rd International Conference on Software Engineering (ICSE)}};
\end{tikzpicture}
\newcommand\copyrightnotice[1]{
    \begin{tikzpicture}[remember picture,overlay]
    \node[anchor=south,yshift=40pt] at (current page.south) {\parbox{\dimexpr 1\textwidth-\fboxsep-\fboxrule\relax}{\footnotesize #1}};
    \end{tikzpicture}
}
\copyrightnotice{
xxxxx~\copyright~2023~IEEE. \\ 
DOI xxxxx
}


\lin{do we have time to add KNOD results?}

\todo{make capital letters in subsection title consistent}

\section{Introduction}
\noindent Automated program repair (APR)~\cite{apr-review,general-apr-1,general-apr-2} helps developers improve software reliability by generating patches automatically to repair software defects. Many deep learning (DL)-based APR techniques~\cite{sequencer,dlfix,coconut,cure,codit,rewardrepair,recoder,knod,transfer-vul,selfapr} adapt DL models to take a buggy software program as input and generate a patched program. 
A typical DL-based APR technique builds a neural network model from a \emph{training set}, which are pairs of buggy code and the corresponding fixed code. Then these models are evaluated on a \emph{test set}, which are also pairs of the buggy and fixed code that is disjoint from the training set.
With the strong learning capability of DL models, these techniques learn diverse patterns of transforming buggy programs to patched programs from large code corpora, and many~\cite{cure, recoder, rewardrepair} outperform traditional template-based~\cite{simfix-template-1,tbar-template-2}, heuristic-based~\cite{arja-heuristic-1,heuristic-2,elixir-heuristic-3} and constraint-based~\cite{ACS-constraint-1,nopol-constraint-2,constraint-3} APR techniques. 

Although DL-based APR techniques are one of the most effective, these tools fail to fix a large portion of bugs. In addition, existing DL-based APR tools typically have to generate hundreds to thousands of candidate patches and take hours to validate these patches to fix enough bugs~\cite{coconut,cure,rewardrepair,recoder}. Recent work shows that 93\% of developers are only willing to review up to ten patches and 63\% of developers expect APR tools to respond within one hour~\cite{trust-apr}. Thus, there is a gap for DL-based APR research to be used in practice~\cite{trust-apr}. 

In addition to DL-based APR models, \emph{code language models} (CLMs)~\cite{plbart,codet5} have shown their promises for fixing bugs, 
given the demonstrated effectiveness of language models~\cite{gpt-1,gpt-2,gpt-3,gpt-j,gpt-neo,bert,roberta,bart,graph-bert,turing-nlg} in natural language domains. 
Different from  DL-based APR models that use an APR-specific design and are trained with labeled APR training sets (typically pairs of buggy and fixed code),  CLMs are trained 
with \emph{huge-sized unlabeled} code corpora (e.g., programs) for general code language modeling tasks, e.g., next token prediction~\cite{codegen,codex}. 

Despite the success of CLMs in many domains~\cite{codet5,codegen,plbart,codebert,graph-codebert}, there has been little \emph{comprehensive, in-depth} work analyzing and comparing CLMs' fixing capabilities in the APR domain with those of existing DL-based APR techniques that are specially designed to fix bugs. Thus, it is an interesting and important research question to ask: 
\textbf{do code language models bring improvement to automated program repair and how?}

\subsection{Evaluation of CLMs on APR Benchmarks}
\label{intro-evaluation}
\noindent Existing techniques~\cite{plbart, codet5,codebert,graph-codebert} evaluate CLMs' fixing capabilities mostly on the benchmark provided by CodeXGLUE~\cite{codexglue}, which is abstracted code (e.g., \code{VAR1 < VAR2.length()}) instead of real-world code (e.g., \code{charno <  sourceExcerpt.length()}).
But understanding and generating concrete variable and function names is a mandatory and challenging step to fix bugs~\cite{cure,coconut,recoder,rewardrepair}. 
In addition, they report BLEU scores (which measure the similarity between generated patched code and the developer patch) instead of validating the correctness of the generated patches. They cannot do so, because CodeXGLUE~\cite{codexglue}  
also contains no full project context or test cases. But many patches with a good BLEU score are incorrect patches. 
Thus, we need to use an APR benchmark with realistic, real-world bugs and test cases to evaluate the true effectiveness of CLMs in fixing bugs. 

While we can and will use real-world APR benchmarks such as \defectsold{}~\cite{defects4j}, \defectsnew{}~\cite{defects4j}, and \quixbugs{}~\cite{quixbugs}, to evaluate the fixing capabilities of CLMs~\cite{codex-quixbugs}, there is a challenge that the training data of these CLMs may contain the bugs or fixes in these APR benchmarks, since these CLMs use public repositories such as all GitHub repositories by a certain date~\cite{codet5,codegen,plbart,codebert,graph-codebert}. While data leaking is a threat to all CLM-related papers, not just this paper, this threat is less of a concern for APR compared to other domains~\cite{barz2020we, tan2015online}, since CLMs do not see the pairs of buggy code and their fixed code during training, and their training data often contains at most the buggy code or the fixed code, but not both. We address this data-leaking challenge by manually creating a new evaluation benchmark \emph{HumanEval-Java} that has not been seen by any of the evaluated CLMs during training. 

In this work, we evaluate ten code language models of four types (PLBART~\cite{plbart}, CodeT5~\cite{codet5}, CodeGen~\cite{codegen}, and InCoder~\cite{incoder}) on four benchmarks (\defectsold{}~\cite{defects4j}, \defectsnew{}~\cite{defects4j}, \quixbugs{}~\cite{quixbugs} and HumanEval-Java). We run developer-written test cases to validate the correctness of generated patches to evaluate and compare the ten CLMs. 

\subsection{Fine-tuning CLMs for the APR Task}
\noindent CLMs are often trained  for  general tasks (e.g., next token prediction) on a large corpus. This training is referred to as \emph{pre-training}. 
\emph{Fine-tuning} is a common technique to train a pre-trained CLM with data from a downstream task, e.g., code summarization or code translation, 
when one wants to apply a general pre-trained CLM to a specific downstream task~\cite{codebert,graph-codebert,codegen,codet5,plbart}. 
Fine-tuning is typically very effective to tailor a general pre-trained CLM for a downstream task~\cite{bert,gpt-1,roberta,bart,T5}. Yet, none of the CLMs have been fine-tuned for the APR task with real-world, non-abstracted APR training data. 

To study how fine-tuning may enhance or hurt CLMs on APR, we fine-tune ten CLMs with APR training data used by DL-based APR techniques.
We also study the impact of the size of the fine-tuning data. Typically the more training data, the more effective the resulting models are up to a certain point, when one trains a DL model from scratch~\cite{data-size-1,T5,gpt-2,gpt-3}. And DL-based APR tools also fix more bugs when they are trained with more data~\cite{recoder}. We will study if fine-tuning with more data improves the fixing capabilities of CLMs for APR~\cite{data-size-1,data-size-2}. 

\subsection{Fixing Capability versus Cost}
\noindent As the size (i.e., number of parameters) of CLMs grows exponentially in recent years~\cite{turing-nlg},  the cost of  applying such large models (e.g., time and memory cost) 
also grows dramatically. Although larger CLMs may fix more bugs, the trade-off between fixing capability and cost of applying such large models is important. Thus, we study as the model sizes change, how the fixing capabilities change  (size efficiency),  how the average time required to generate patches changes (time efficiency), and how the memory footprint changes (memory efficiency) of the ten different-sized CLMs. 

\subsection{Contributions}
\noindent To sum up, this paper makes the following contributions:

\smallskip \noindent (1) \ul{A new APR benchmark}, HumanEval-Java, that no existing CLMs have seen during pre-training, to ensure the fairness of evaluation.

\smallskip \noindent (2) \ul{A study of ten CLMs} (of four  architectures, i.e., PLBART, CodeT5, CodeGen, and InCoder) and \ul{four state-of-the-art DL-based APR techniques} (i.e., CURE, RewardRepair, Recoder, and KNOD) \ul{on four APR benchmarks} (i.e., \defectsold{}, \defectsnew{}, \quixbugs{} and our new HumanEval-Java):
\begin{itemize}
    \item \textbf{Finding 1:} CLMs, even without fine-tuning, have competitive fixing capabilities. To our surprise, the best CLM, as is, fixes 72\% more bugs than the state-of-the-art DL-based APR techniques.
    \item \textbf{Finding 2:} While buggy lines (code lines that need to be modified) are useful to guide fixing, CLMs fail to make good use of them and fix fewer bugs when the buggy lines are explicitly given.
\end{itemize}

\smallskip \noindent (3) \ul{The first fine-tuning experiment} of ten CLMs with APR training data, and \ul{a study of fine-tuning's impact} (and its training data size) on CLMs for APR:
\begin{itemize} 
    \item \textbf{Finding 3:} Fine-tuning improves CLMs' fixing capabilities by 31\%--1,267\%, and fine-tuned CLMs outperform DL-based APR techniques significantly, by 46\%--164\%.
    \item \textbf{Finding 4:} Although fine-tuning helps, it sometimes makes CLMs over-rely on the buggy lines, and thus fail to fix some bugs that can be fixed without fine-tuning.
    \item \textbf{Finding 5:} CodeT5 and CodeGen models achieve the best fixing capabilities after being fine-tuned with 10,000 APR training instances, and fine-tuning with more APR data makes them fix fewer (reduced by 8\%--19\%). InCoder model fixes the most bugs after being fine-tuned with 50,000 training instances, and more fine-tuning data also makes it fix fewer (reduced by 9\%).
\end{itemize}

\smallskip \noindent (4) \ul{A study of the size, time and memory efficiency} of CLMs (i.e., PLBART, CodeT5, CodeGen and InCoder).
\begin{itemize}
    \item \textbf{Finding 6:} CodeT5 and InCoder models have the best size efficiency, suggesting developing larger CodeT5 or InCoder models is the most promising. Besides, CodeT5, PLBART, and InCoder models are better choices given limited resources, as they have better time and memory efficiency than CodeGen models.
\end{itemize}

\smallskip \noindent (5) \ul{Implications for future work} 
(Section~\ref{implication}).

\section{Code Language Models}

\subsection{CLM Architectures}
\label{language model architecture}
\noindent Code language models can be categorized into three groups: \emph{encoder-only} models, \emph{decoder-only} models, and \emph{encoder-decoder} models. 
Regardless of groups, most existing code language models are built with a Transformer~\cite{transformer} architecture as it has the best learning capability and the greatest scalability~\cite{transformer,bert,roberta,bart,T5,gpt-1,gpt-2,gpt-3}. 

Encoder-only models include~\codebert{}~\cite{codebert} and~\graphcodebert{}~\cite{graph-codebert}, which only have a bidirectional transformer encoder~\cite{transformer} with attention mechanism~\cite{transformer} to learn vectorized embedding of the input code sequence. As they only have encoders, these models are most suitable for downstream tasks that require no generation, such as code representation (i.e., embedding the input code) and code clone detection~\cite{codebert}.

Decoder-only models include~\codegen{}~\cite{codegen}, InCoder~\cite{incoder}, and~\codex{} ~\cite{codex}, which only have an autoregressive transformer decoder~\cite{transformer} to learn to generate code sequences. Different from encoder-only models that calculate embedding for input code, decoder-only models are most suitable for downstream tasks such as open-ending code generation~\cite{codex}, i.e., generating code based on  input prompts, where a prompt could be natural language text describing the functionality, the signature of a function, or the first few lines of code in a function. 

Finally, encoder-decoder models include~\plbart{}~\cite{plbart} and~\codetf{}~\cite{codet5}, which have both a bidirectional transformer encoder and an autoregressive transformer decode. The encoder is trained to calculate the embedding of input code and the decoder is trained to generate code. Thus, encoder-decoder models are more flexible to suit both non-generation (e.g., code classification)~\cite{plbart} and generation (e.g., code summarization)~\cite{codet5} downstream tasks.

\begin{figure}[htb]
    \centering
    \includegraphics[width=0.488\textwidth]{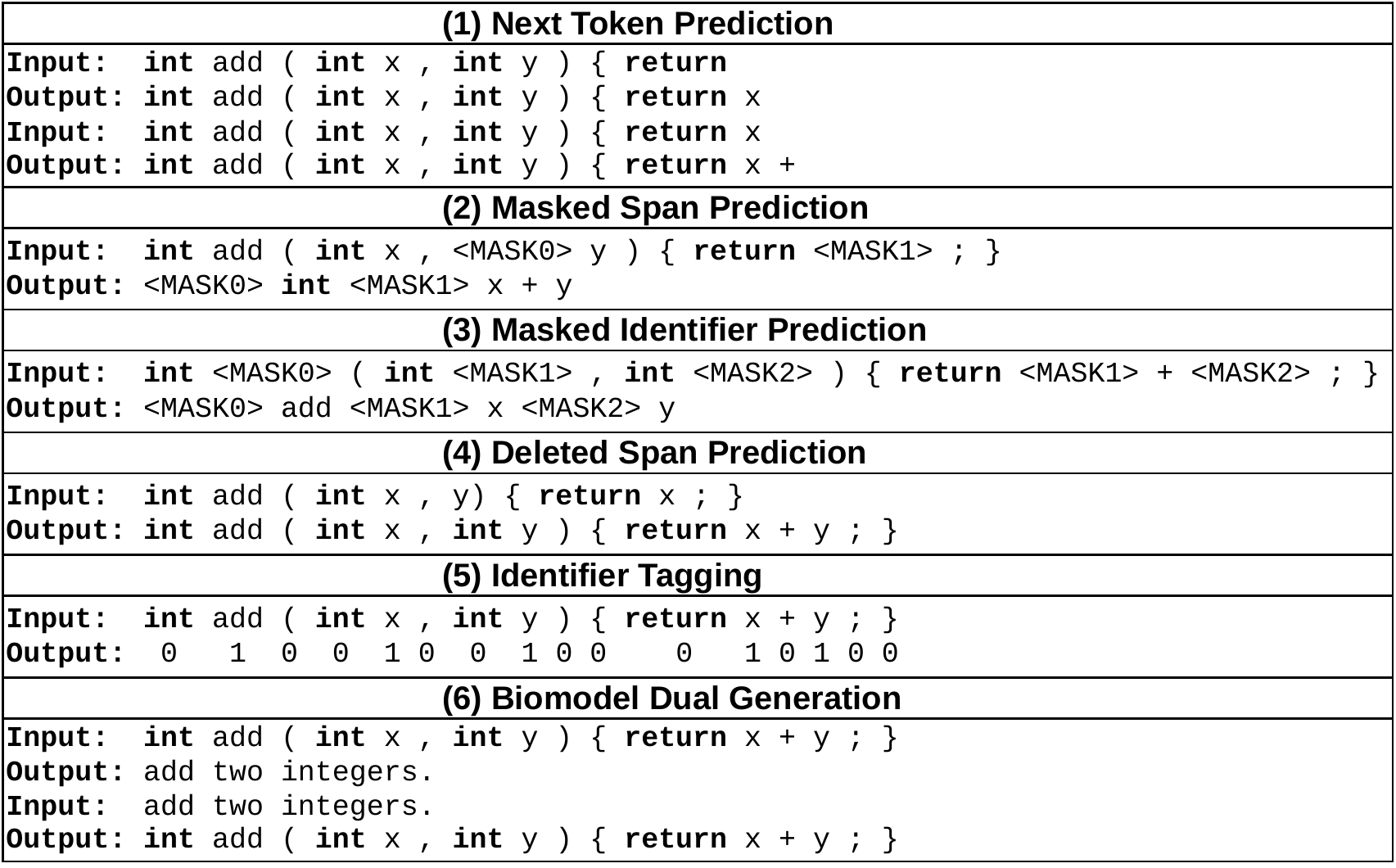}
    \caption{Common pre-training tasks with examples that are used by existing CLMs.
    }
    \label{fig:pretraining task}
\end{figure}

\subsection{Common Pre-training Tasks}
\label{pre-training task}
\noindent An important design difference among code language models is the tasks used in pre-training. Figure~\ref{fig:pretraining task} shows  several common pre-training tasks used by existing code language models.

\smallskip \noindent \textbf{(1) Next Token Prediction} is the task that given a piece of incomplete code, the language model is trained to predict the following  token, e.g., given \code{int add(int x, int y)\{ return}, the next token should be \code{x}. This process, if done iteratively, trains language models to generate complete programs from the beginning to the end. 
\lin{do you mean complete methods, programs? }\nan{both are doable}

\smallskip \noindent \textbf{(2) Masked Span Prediction} is the task of training a language model to predict the masked code snippet in the input code. Figure~\ref{fig:pretraining task} shows a simple Java function that returns the sum of two integers, where some parts of the function are masked by placeholders \code{<MASK0>} and \code{<MASK1>}. The language model is trained to predict that \code{<MASK0>} should be \code{int} and \code{<MASK1>} should be \code{x + y}.

\smallskip \noindent \textbf{(2) Masked Identifier Prediction} is the task of predicting identifier names in the given code. For example, all three identifiers (\code{add}, \code{x}, and \code{y}) are masked by placeholders (\code{<MASK0>}, \code{<MASK1>}, and  \code{<MASK2>}), and a code language model is trained to predict their correct name. 

\smallskip \noindent \textbf{(3) Deleted Span Prediction} is the task that given code with some parts deleted (e.g., \code{int} and \code{+ y} are deleted), the language model is trained to generate the completed code.

\smallskip \noindent \textbf{(4) Identifier Tagging} is the task of predicting whether each token in the code is an identifier or not. For example, \code{add} is an identifier while \code{int} is not. Thus the predicted label of \code{add} is 1 and that of \code{int} is 0.

\smallskip \noindent \textbf{(5) Biomodel Dual Generation} is the task of training a language model that transforms  code between different languages, e.g., from a natural language description to Java code, from Java code to natural language text, or from Java code to Python code, etc.

\section{Experimental Design}
\begin{figure}[htb]
    \centering
    \includegraphics[width=0.48\textwidth]{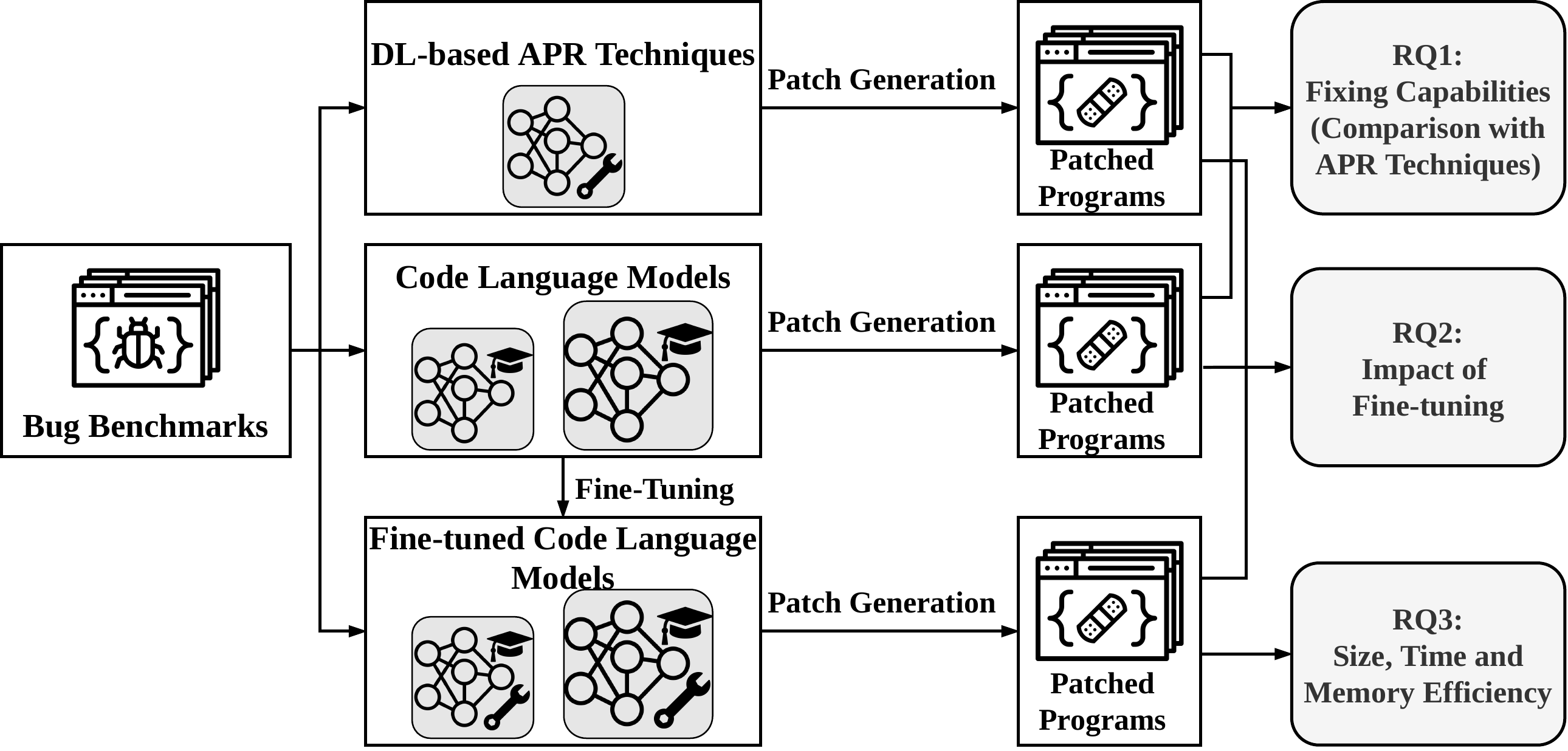}
    \caption{Overview of experimental design.}
    \label{fig:overview}
\end{figure}

\noindent Figure~\ref{fig:overview} shows the overview of our experimental design. We apply ten CLMs and three state-of-the-art DL-based APR techniques on four bug benchmarks---three widely-used and one new benchmark that we designed to address the data-leaking challenge---to generate patches. We study and compare the fixing capabilities of the CLMs and the DL-based APR techniques (RQ1). Then we fine-tune these CLMs with APR training datasets of different sizes, and study the patches generated by fine-tuned CLMs to show the impact of training data size on fine-tuning (RQ2). 
Finally, by comparing patches generated by differently sized code language models, we study the size, time, and memory efficiency of different CLMs (RQ3).

We focus on Java single-hunk bugs, as the best DL-based APR techniques are all specially designed for Java single-hunk bugs (i.e., the buggy code is continuous lines of code)~\cite{cure, recoder, rewardrepair}. This enables us to explore how CLMs are different from DL-based APR techniques in fixing \emph{the same types of bugs}.

\subsection{Three Existing Bug Benchmarks}
\noindent 
We select real-world bug benchmarks widely used in the APR domain, including\textbf{~\defectsold{}}~\cite{defects4j},\textbf{~\defectsnew{}}~\cite{defects4j}, and\textbf{~\quixbugs{}}~\cite{quixbugs}. \defectsold{} is the most widely used version of the Defects4J benchmark that contains 393 bugs, among which 130 are single-hunk bugs. \defectsnew{} is the latest version of the Defects4J benchmark that contains 444 additional bugs, among which 108 are single-hunk bugs. Both \defectsold{} and \defectsnew{} are collected from famous Java projects such as ``Google Closure compiler'' and ``Apache commons-math''. \quixbugs{} is a benchmark that contains 40 bugs regarding famous algorithms like ``quick sort'' and ``merge sort''. 

\subsection{New HumanEval-Java Benchmark to Avoid Data Leaking}
\noindent \defectsold{}, \defectsnew{}, and \quixbugs{} are widely used by DL-based APR techniques. However, applying CLMs only on them might be problematic, as CLMs may have seen these benchmarks in their pre-training data. By checking CodeSearchNet~\cite{codesearchnet} and BigQuery\footnote{https://console.cloud.google.com/marketplace/details/github/github-repos?pli=1}, which are the data sources of common CLMs, we find that four repositories used by the \defects{} benchmark are also in CodeSearchNet, and the whole \defects{} repository is included by BigQuery.
Thus, it is very likely that existing APR benchmarks are seen by CLMs during pre-training. \nan{revised}

To overcome this threat, we create a new bug benchmark from HumanEval~\cite{codex}, named \textbf{HumanEval-Java}. HumanEval is a dataset manually created to address the threat that CLMs may have seen test datasets available online. Yet, it is created for evaluating code generation task~\cite{codex} and is written in Python. We manually convert the Python programs in HumanEval and their test cases into Java programs and Junit test cases, and then inject bugs in the correct Java programs to create an APR benchmark. HumanEval-Java contains 164 (single-hunk) Java bugs, varying from simple bugs like incorrect operator usage to complex logical bugs that require modification of several lines of code to fix. Since HumanEval-Java is converted from HumanEval and the bugs are manually injected, none of the CLMs would have seen it before. Thus, it is the fairest benchmark to compare CLMs with DL-based APR tools.

\subsection{Studied Code Language Models}
\noindent Table~\ref{tab:language models} lists the ten CLMs evaluated in this paper. We select CLMs to study based on the following requirements: (1) the language model is trained on a large enough code corpus (e.g., we exclude T5~\cite{T5} and GPT-2~\cite{gpt-2}, which are natural language models, and  we exclude GPT-Neo~\cite{gpt-neo} and GPT-J~\cite{gpt-j}, which are  trained on the THEPILE dataset~\cite{thepile}, of which 90\% are English text), (2) the language model can be applied to APR without any modification to its  architecture or extra designs. Thus, encoder-only models such as~\codebert{}~\cite{codebert} or~\graphcodebert{}~\cite{graph-codebert} are excluded. They need either an extra decoder or careful designs of input format to be applied to generate patches\nan{added more why CodeBERT is excluded},
and (3) the pre-trained language model is publicly accessible  (e.g., \codex{}~\cite{codex} is excluded as its model is not released and cannot be fine-tuned). As a result, we select four types of code language models, which are PLBART~\cite{plbart}, CodeT5~\cite{codet5}, CodeGen~\cite{codegen}, and InCoder~\cite{incoder}.

\begin{table}[]
    \centering
    \scriptsize
    \begin{tabular}{l@{\hspace{3pt}}l|
    r@{\hspace{4pt}}|
    r@{\hspace{4pt}}|r|r}
    \hline
         & & \textbf{PLBART} & \textbf{CodeT5} & \textbf{CodeGen} & \textbf{InCoder} \\
    \hline
    \hline
        \multirow{3}{*}{Models} & & base (140M) & small (60M) & 350M & 1B \\
        & & large (400M) & base (220M) & 2B & 6B \\
        & & & large (770M) & 6B & \\
    \hline
       \multicolumn{2}{l|}{Data Source} & StackOverflow & CodeSearchNet & THEPILE & StackOverflow \\
        & & BigQuery & BigQuery & BigQuery & GitHub/GitLab \\
    \hline
        \multirow{2}{*}{Raw Size} & NL & 79.0GB & - & 1.1TB & 57.0GB\\
        & PL & 576.0GB & - & 436.3GB & 159.0GB \\
        \multirow{2}{*}{Instances} & NL & 47M & 5M & - & - \\
        & PL & 680M & 8M & - & - \\
        \multirow{2}{*}{Tokens} & NL & 6.7B & - & 354.7B & - \\
        & PL & 64.4B & - & 150.8B & - \\
    \hline 
    \end{tabular}
    \caption{Ten CLMs of four architectures that we study in this work. NL refers to natural language, while PL refers to programming language. ``Raw Size'' refers to the size of collected pre-training data, ``Instances'' refers to the number of pre-training data fed to the models, and ``Tokens'' is the total number of NL or PL tokens in the pre-training data. ``-'' means the corresponding number is not reported.
   }
    \label{tab:language models}
\end{table}

\begin{figure*}[ht]
    \centering
    \includegraphics[width=0.98\textwidth]{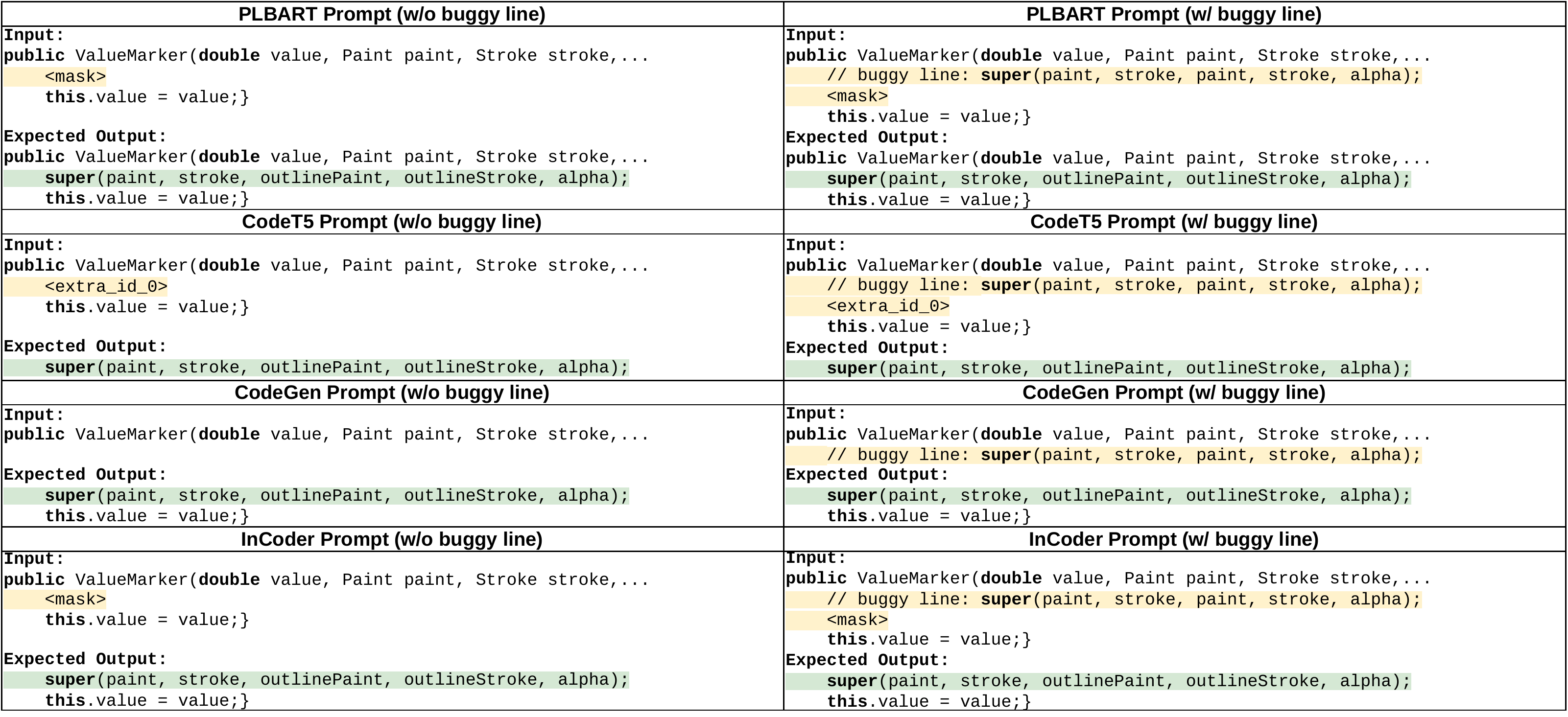}
    \caption{Prompts used for applying  CLMs on the APR task, using Chart-20 in~\defectsold{} benchmark as an example.}
    \label{fig:prompt}
\end{figure*}

\smallskip\noindent \textbf{PLBART:} These models follow BART's~\cite{bart} encoder-decoder architecture (Section \ref{language model architecture}), and are pre-trained on masked span prediction and deleted span prediction (Section \ref{pre-training task}).

The pre-training dataset of PLBART comes from BigQuery and StackOverflow, which contains 47M natural language instances and 680M programming language instances. The programming language data consist of 470M Java instances (69\%) and 210M Python instances (31\%). Thus, it performs better on Java and Python code than other programming languages~\cite{plbart}.

The developers released two PLBART models of different sizes, referred to as \textbf{PLBART-base} (140M parameters) and \textbf{PLBART-large} (400M parameters), both of which are pre-trained with the same data and pre-training tasks. We include both models in this work.

\smallskip\noindent \textbf{CodeT5:} The CodeT5 models follow T5's~\cite{T5} encoder-decoder architectures (Section~\ref{language model architecture}), and are pre-trained on several code-specified tasks, including masked span prediction, identifier tagging, masked identifier prediction, and bimodal dual generation (Section~\ref{pre-training task}). 

The pre-training dataset comes from CodeSearchNet~\cite{codesearchnet} (an open-sourced code corpus containing 4291 projects collected from GitHub), and also C and CSharp programs collected via BigQuery, which contains around 5M natural language instances and 8M programming language instances. The programming language data consists of 22\% JavaScript, 18\% Java, 13\% Python, 13\% CSharp, 12\% C, 11\% PHP, 8\% Go and 2\% Ruby languages.

Developers released three CodeT5 models of different sizes, referred to as \textbf{CodeT5-small} (60M parameters), \textbf{CodeT5-base} (220M parameters), and \textbf{CodeT5-large} (770M parameters). We include all three models in this work.

\smallskip\noindent \textbf{CodeGen:} The CodeGen models follow a  decoder-only architecture (Section~\ref{language model architecture}) and are pre-trained on the next token prediction task (Section~\ref{pre-training task}). The developers  released 12 CodeGen models: CodeGen-350M/2B/6B/16B-NL/Multi/Mono~\cite{codegen}. The ``350M/2B/6B/16B'' in the name refers to the number of parameters, while ``NL/Multi/Mono'' specifies different training datasets. Specifically, ``NL'' denotes that the model is only pre-trained on THEPILE~\cite{thepile} dataset (mostly English text). ``Multi'' means that the model is also pre-trained on data collected via BigQuery, which includes 30\% Java, 21\% C++, 17\% C, 16\% Python, 8\% JavaScript and 8\% Go languages. ``Mono'' specifies that the model is also pre-trained on a huge Python corpus collected from GitHub. 

As we focus on fixing Java bugs in this paper, and the 16B model is too large to run on our machines, we only study the CodeGen-350M/2B/6B-Multi models, referred to as \textbf{CodeGen-350M}, \textbf{CodeGen-2B}, and \textbf{CodeGen-6B} to simplify the names. \nan{revised}

\smallskip\noindent \textbf{InCoder:} The InCoder models follow XGLM~\cite{xglm}'s decoder-only architecture (Section \ref{language model architecture}) and are also pre-trained on the masked span prediction task (Section \ref{pre-training task}). 

The pre-training data of InCoder comes from open-sourced projects on GitHub and GitLab, and StackOverflow posts, which consists of 159 GB of code data (33\% Python, 26\% JavaScript, 10\% C/C++, 8\% HTML, 4\% Java, etc.) and 57 GB of text data (the number of instances and tokens is not reported). The pre-training data is deduplicated and filtered to guarantee high quality.

Developers released two InCoder models of different sizes, referred to as \textbf{InCoder-1B} (1B parameters) and \textbf{InCoder-6B} (6B parameters), both of which are included in this work.

\subsection{Applying Code Language Models}
\noindent To answer RQ1, we apply the pre-trained CLMs without any fine-tuning to study their fixing capabilities learned from pre-training tasks. We carefully check their papers and documentation to ensure we set them up correctly. 

To set up a fair comparison, we apply each code language model with two different prompts, i.e., the input to a CLM. The first prompt does not contain the buggy lines (but the bug location is still known), which is the natural (default) way of applying these CLMs according to their documentation. The second prompt gives the buggy lines as lines of comments, to ensure CLMs have the same information as DL-based APR techniques, which require buggy lines and surrounding functions to fix bugs. Figure~\ref{fig:prompt} shows the prompts for different CLMs. 

\begin{itemize}
    \item To apply PLBART models without providing the buggy lines, the whole buggy function is provided, with the buggy lines masked by a placeholder (specifically, \code{$<$mask$>$} for PLBART). The models are expected to output the \emph{whole patched function}. To validate the correctness, the test cases are executed on the output function. \lin{isn't a harder task to generate the whole method compared to CodeT5 as it may change the context/correct code? You may present CodeT5 first, so that you can explain this difference when introducing PLBART}\nan{I agree it's harder. I would explain below but prefer not to change the order. The whole paper goes as PLBART-CodeT5-CodeGen}
    \item To apply CodeT5 models without the buggy lines, the input format is the same as PLBART, but the placeholder used to mask the buggy line is \code{$<$extra\_id\_0$>$}. CodeT5 models are expected to generate the patched lines. Since CodeT5 models do not have to generate the whole function, it is supposed to be easier for CodeT5 to generate the correct patch. \nan{explained the difference here.}
    To validate the correctness of the patch, we replace the buggy lines with CodeT5's output to form the patched program, on which developer-written test cases are executed.
    \item To apply CodeGen models without the buggy lines, the input is the function before the buggy lines. The models are expected to complete the function by generating the patched line and the remainder of the function after the buggy lines. To validate the correctness, we append the output to the input to form a complete function, on which the test cases are executed. CodeGen models do not know the code snippet after the buggy lines (thus, they have less information when fixing bugs), which is due to the design of CodeGen.
    \item To apply InCoder models without the buggy line, the input is the same as PLBART. The models are expected to output the patched line as well as the code after the patched line. To validate the correctness, we append the output to the code before the buggy line to form a complete function and run the test cases. Compared with CodeGen models, InCoder models have the code snippet after the buggy lines when generating patches, which is the advantage of the design of InCoder models (i.e., using masked span prediction as a pre-training task).
    \item To apply these CLMs models with buggy lines as part of the prompts, the buggy lines are provided as lines of comments before the location of the buggy lines. 
\end{itemize} 

\begin{figure}[t]
    \centering
    \includegraphics[width=0.48\textwidth]{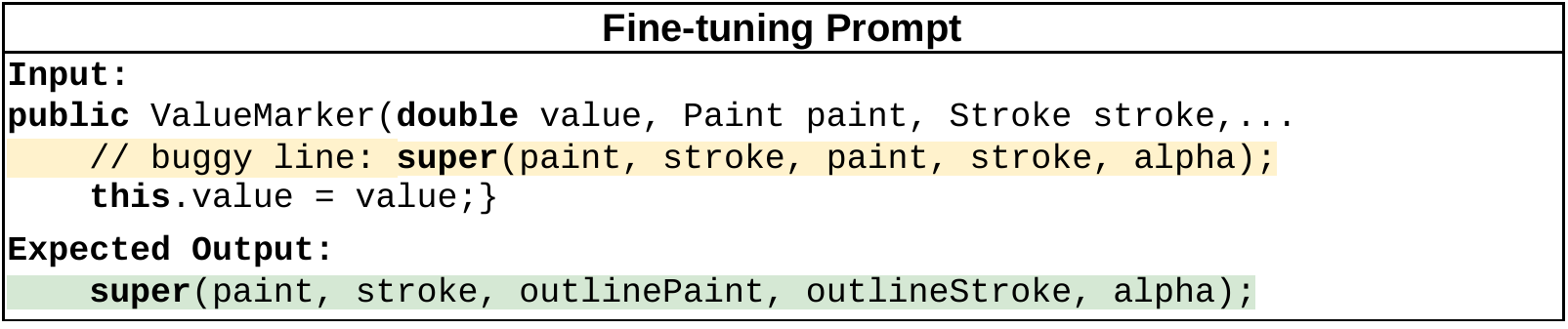}
    \caption{Prompt used for fine-tuning CLMs with Chart-20 in~\defectsold{} benchmark as an example.}
    \label{fig:finetune prompt}
\end{figure}

\subsection{Fine-tuning Code Language Models}
\noindent To answer RQ2, we conduct the first experiment to fine-tune ten CLMs for the APR task. 
Since the APR training data are pairs of buggy code and fixed code, we provide the buggy code as part of the prompts to CLMs so that they can learn to generate fixed code given buggy code (Figure~\ref{fig:finetune prompt}). 
Similar to RQ1, to make the comparison among fine-tuned code language models fair, we use the same prompt, i.e., input to the models, for all language models. 
\lin{removed fine-tuning and inference. you meant the same prompt for all models, but as is, it is ambiguous as it may mean the same prompt for fine-tuning and inference. or need to rephrase.} 
We fine-tune the CLMs to directly output the patched lines, the same as the output of DL-based APR techniques. 
 
For the training data for fine-tuning, we use the APR data shared in previous work~\cite{recoder}, which is collected from commits  of open-sourced GitHub Java projects, and treat each single-hunk fix as a separate instance, which contains 143,666 instances in total. The dataset is randomly split into a training dataset with 129,300 instances and a validation dataset with 14,366 instances to tune the hyper-parameters (e.g., number of training epochs).

For all the CLMs, we apply the same setting for  fine-tuning. Specifically, the batch size is one (due to our hardware constraints). We use the Adam optimizer~\cite{huggingface} with a learning rate of $1e^{-5}$ to update the model weights. CLMs are only fine-tuned for one epoch over the APR dataset, as they converge fast on the validation dataset. We set a fixed random seed when fine-tuning different models to minimize variance for a consistent, fair comparison.

\subsection{Baseline DL-based APR Techniques}
\noindent To compare CLMs with APR tools, we select the four best open-sourced DL-based APR techniques, namely \cure{}~\cite{cure}, \rewardrepair{}~\cite{rewardrepair}, \recoder{}~\cite{recoder}, and KNOD~\cite{knod}. Other APR techniques either fix fewer bugs~\cite{coconut,tbar-template-2,dlfix,sequencer,simfix-template-1} or are unavailable~\cite{alpharepair}, and
we need to select open-sourced techniques to apply them to our new benchmark HumanEval-Java.
These APR techniques all have encoder-decoder architecture, but also have APR-specific designs.

\cure{} implements its encoder and decoder with convolutional networks~\cite{fairseq}, and applies a small code GPT~\cite{cure,gpt-1} to learn code syntax (but it is only pre-trained on 4M Java functions), and designs a code-aware search strategy to exclude invalid identifiers during patch generation~\cite{cure}. 

\rewardrepair{} is implemented with transformer architecture and is the most similar to CLMs regarding architectures. It also considers patch execution information (compilability and correctness) in the calculation of loss function during training, which makes the model learn to generate compilable and correct patches~\cite{rewardrepair}. 

\recoder{} has a novel architecture to generate edits to modify the abstract syntax tree (AST) to patched AST~\cite{recoder}. Generating at the AST level enables it to generate more syntactically correct patches, and generating edits enables it to fix bugs with fewer decoding steps. 

\rev{KNOD is a recent DL-based APR technique that uses graph-transformer~\cite{graph-transfrmer-1} and a novel three-stage tree decoder to generate patched ASTs. It also uses domain-knowledge distillation~\cite{firstlogic} to help the model learn code syntaxes and semantics.}

\subsection{Patch Generation and Validation}
\noindent For all experiments, we let each tool (CLMs, fine-tuned CLMs, or DL-based APR techniques) generate ten candidate patches for each bug and run the developer-written test cases on the patched program. The first patched program that passes all the test cases is considered a plausible patch. And we finally manually check the correctness of plausible patches to distinguish correct patches (which should be identical or semantically equivalent to developer-written patches).

\begin{table*}[ht]
    \centering
    \begin{tabular}{l|r|rr|rrr|rrr|rr|rrrr}
    \hline
        \textbf{Benchmarks} & \textbf{\#Bugs} &  
        \multicolumn{2}{c}{\textbf{PLBART}} &
        \multicolumn{3}{|c|}{\textbf{CodeT5}} &
        \multicolumn{3}{c|}{\textbf{CodeGen}} &
        \multicolumn{2}{c|}{\textbf{InCoder}} &
        \multicolumn{4}{c}{\textbf{DL-based APR Techniques}}\\
        & & base & large & small & base & large & 350M & 2B & 6B & 1B & 6B & CURE & Reward & Recoder & KNOD \\
        \hline
        \hline
        \defectsold{} & 130 & 13 & 13 & 1 & 0 & 1 & 4 & 11 & 11 & 10 & 16 & 6 & 20 & \textbf{24} & 20 \\
        \defectsnew{} & 108 & 9 & 8 & 2 & 4 & 1 & 3 & 4 & 8 & 10 & \textbf{15} & 6 & 8 & 11 & 13\\
        \quixbugs{} & 40 & 11 & 12 & 3 & 0 & 3 & 7 & 15 & \textbf{16} & 14 & 15 & 5 & 7 & 6 & 10\\
        HumanEval-Java & 164 & 39 & 52 & 3 & 5 & 6 & 30 & 49 & 46 & 40 & \textbf{59} & 18 & 22 & 11 & 18\\
    \hline
    \hline
        Total & 442 & 72 & 85 & 9 & 9 & 11 & 44 & 79 & 81 & 74 & \textbf{105} & 35 & 57 & 52 & 61\\
    \hline
    \end{tabular}
    \caption{Number of correct fixes generated by the ten CLMs (without fine-tuning), and four DL-based APR techniques. Reward denotes RewardRepair.
    }
    \label{tab:lm fixing capability}
\end{table*}

\section{RQ1: Fixing Capabilities}
\noindent 
Table~\ref{tab:lm fixing capability} shows the fixing capabilities of the ten CLMs and three state-of-the-art DL-based APR techniques on four bug benchmarks, including our new HumanEval-Java.  
We report the number of correct patches within the \emph{top ten} patches generated by each technique since recent work shows that 93\% of developers are only willing to review up to ten patches~\cite{trust-apr}. The results of CLMs are obtained and reported without feeding buggy lines, as CLMs fix more bugs without buggy lines information (analyzed in Section~\ref{impact of buggy line}).

\subsection{Comparison between CLMs and DL-based APR Techniques}
\noindent 
Table~\ref{tab:lm fixing capability} shows that different types of CLMs perform significantly differently when applied to APR without fine-tuning. In general, PLBART models, CodeGen models (except CodeGen-350M), and InCoder models fix more bugs than APR tools in the four APR benchmarks combined (Row `Total'), while CodeT5 models fix the fewest. Specifically, InCoder-6B fixes the most number (105) of bugs, which is 72\% more than the best DL-based APR technique, KNOD. The second best is PLBART-large, which fixes 85 bugs and is 39\% more than KNOD. The poor result of CodeT5 models might be due to that it is pre-trained for significantly different tasks~\cite{codet5}, including code-to-code generation, code-to-identifier-tag prediction, and code-to-natural-language generation. Thus, without fine-tuning them for the APR task, CodeT5 models cannot generate reasonable code or correct patches. 

\begin{figure}[ht]
    \centering
    \includegraphics[width=0.4\textwidth]{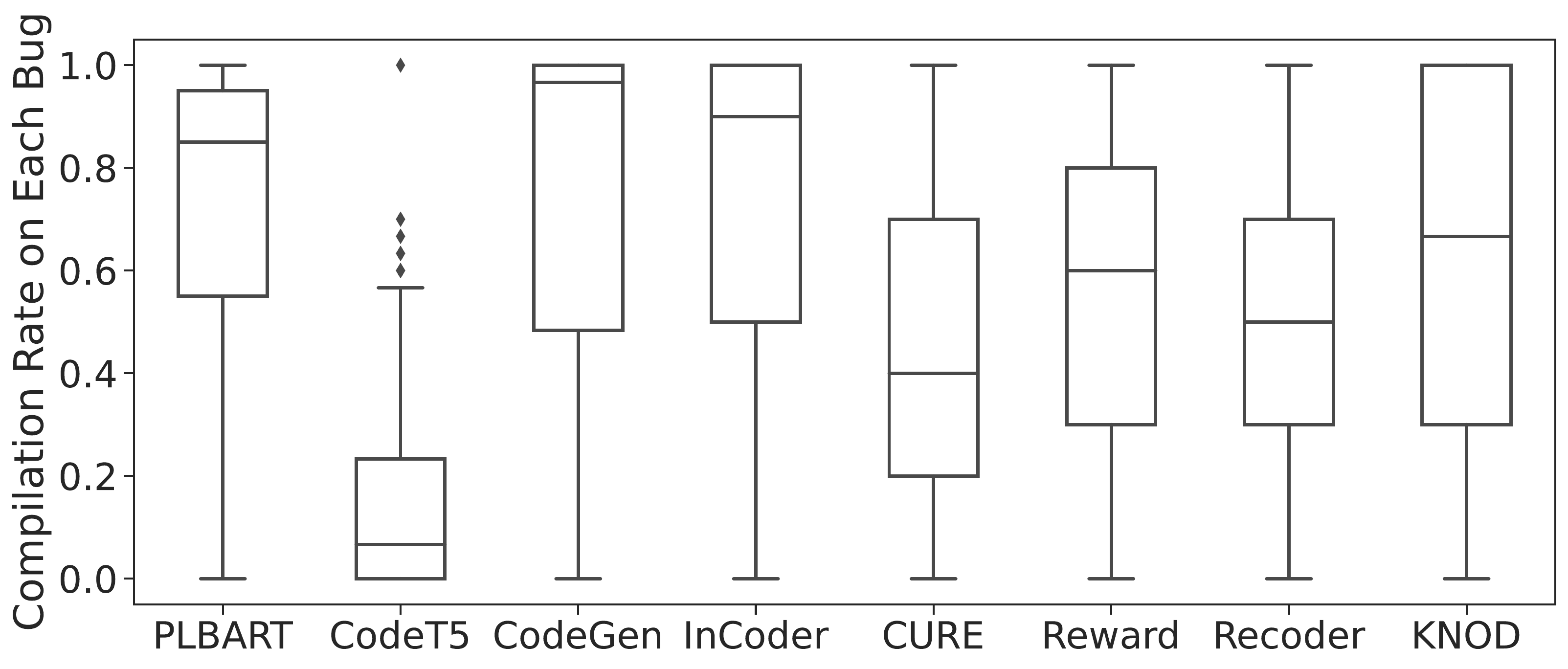}
    \caption{Distributions of compilation rates of the ten patches generated for each bug from all four benchmarks.}
    \label{fig: compilable}
\end{figure}

Figure~\ref{fig: compilable} shows the distributions of the compilation rate of patches generated for  bugs in all benchmarks by each model. CodeGen models generate the most compilable patches, with an average compilation rate of 73\% and a median of 97\%. PLBART and InCoder models also generate much more compilable patches than DL-based APR techniques. DL-based APR techniques are only able to generate 44\%--62\% compilable patches on average. 

\begin{figure}[ht]
    \centering
    \subfigure[Math-75 bug in \defectsold{} that DL-based APR techniques fix but CLMs do not.]{
        \includegraphics[width=0.49\textwidth]{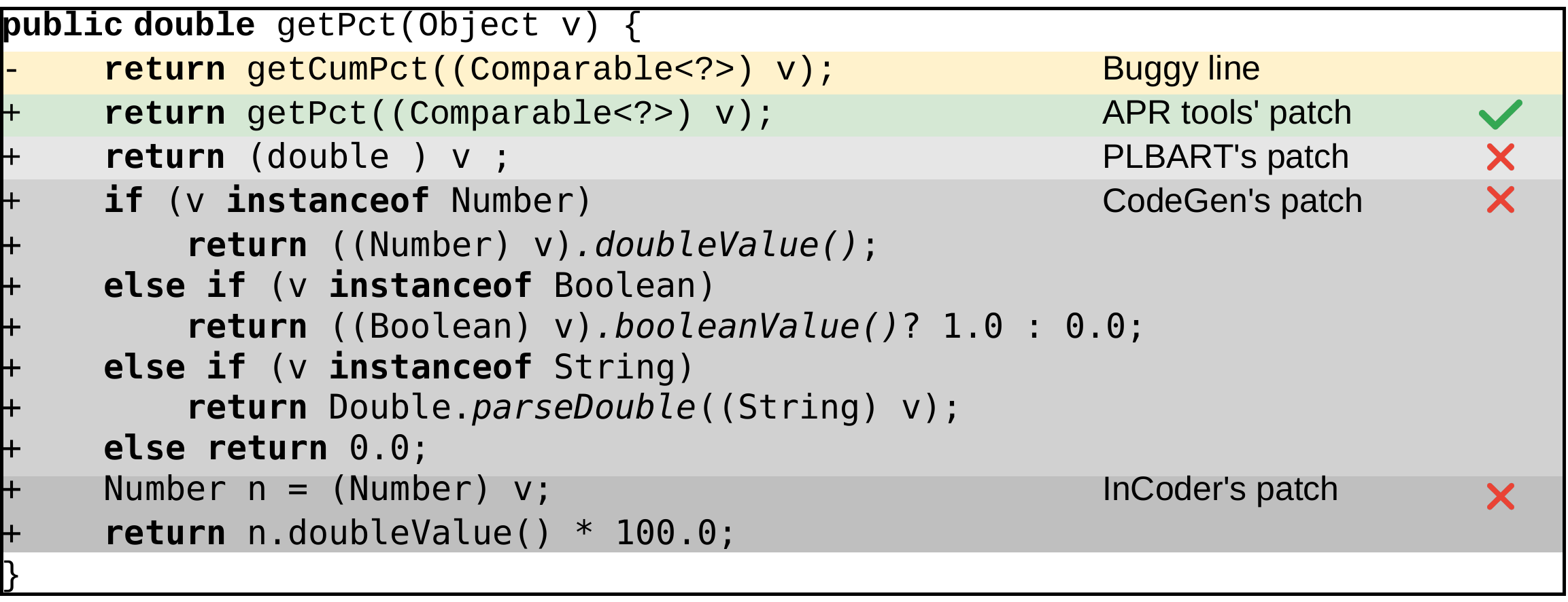}
    }
    \subfigure[GCD in \quixbugs{} that CLMs fix but DL-based APR techniques do not.]{
        \includegraphics[width=0.49\textwidth]{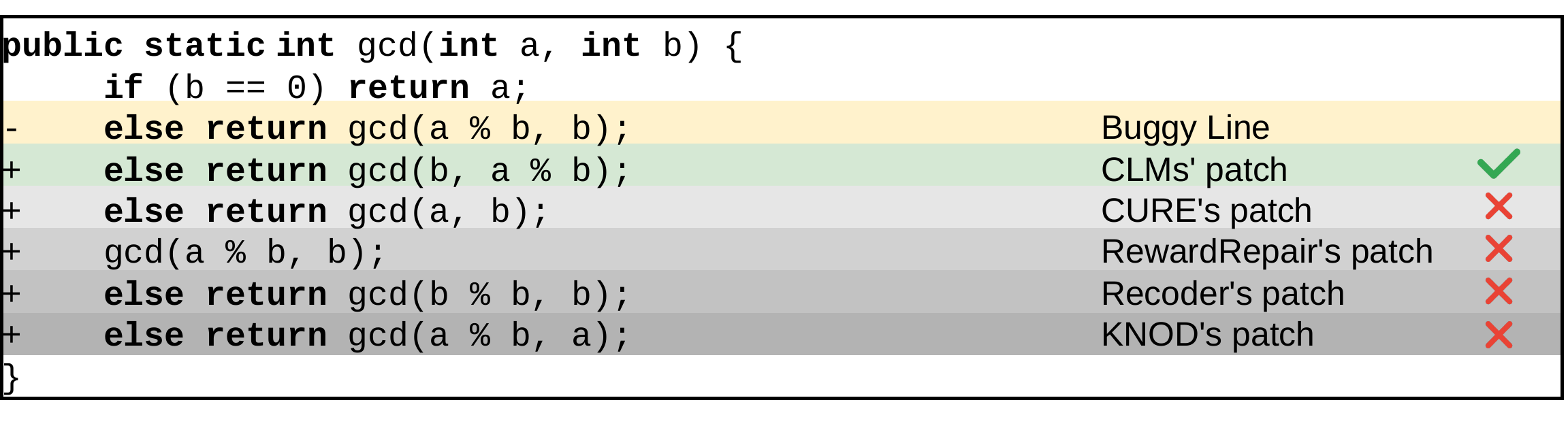}
    }
    \caption{Examples of bugs on which CLMs and DL-based APR tools perform differently.}
    \label{fig:case apr lm}
\end{figure}

CLMs and DL-based APR techniques have different fixing capabilities on different benchmarks. 
Specifically, DL-based APR techniques have better fixing capabilities than CLMs on \defectsold{}. Figure~\ref{fig:case apr lm}(a) shows an example (Math-75) in \defectsold{} that all three DL-based APR techniques fix but all ten CLMs fail to fix. 
This bug has little context. Without enough context, although CLMs can generate reasonable code, they fail to generate the correct fix.
Regardless of providing the buggy line \code{return getCumPct((Comparable<?>) v);} to the CLMs or not, all CLMs fail to fix this bug. As shown later in Section~\ref{impact of buggy line}, CLMs without fine-tuning make poor use of the buggy lines.
In contrast, APR tools are designed to leverage the buggy line information and thus fix this bug.

In contrast, on \quixbugs{} and HumanEval-Java benchmarks, PLBART, CodeGen and InCoder models show much better fixing capabilities than APR tools. Figure~\ref{fig:case apr lm} (b) shows an example (GCD) in \quixbugs{} that PLBART, CodeGen and InCoder models can fix but APR tools cannot. Although CLMs do not see the buggy line \code{else return gcd(a \% b, b);}, they have learned from natural language text or code corpus that \code{gcd} stands for greatest common divisor, and can complete the function correctly. By contrast, APR tools rely on the buggy line a lot when generating candidate patches. Their patches look like applying simple edit operations on the buggy line without considering code syntax and semantics carefully. i.e., CURE's patch replaces \code{a \% b} with \code{a}, RewardRepair's patch deletes \code{return}, which even makes the function uncompilable, Recoder's patch replaces \code{a} with \code{b}, and KNOD's patch replaces \code{b} with \code{a}. 

\finding{CLMs have competitive fixing capabilities even without fine-tuning. PLBART, CodeGen, and InCoder models fix more bugs and generate more compilable patches than state-of-the-art DL-based APR techniques, while CodeT5 models, as an exception, generate poor patches before fine-tuning.}

\subsection{Impact of Buggy Lines}
\label{impact of buggy line}

\begin{table*}[ht]
    \centering
    \begin{tabular}{cc|ccc|ccc|cc}
    \hline
        \multicolumn{2}{c}{\textbf{PLBART}} &
        \multicolumn{3}{|c}{\textbf{CodeT5}} &
        \multicolumn{3}{|c}{\textbf{CodeGen}} &
        \multicolumn{2}{|c}{\textbf{InCoder}} \\
        base & large & small & base & large & 350M & 2B & 6B & 1B & 6B \\
    \hline
        36 (-36) & 62 (-23) & 5 (-4) & 2 (-7) & 4 (-7) & 28 (-16) & 58 (-21) & 73 (-8) & 61 (-13) & 99 (-6)\\
    \hline
    \end{tabular}
    \caption{Number of correct fixes generated by CLMs with  buggy lines provided. Numbers in () are the reduction compared to those without buggy lines.}
    \label{tab:buggyline lm fixing capability}
\end{table*}

\noindent Table~\ref{tab:buggyline lm fixing capability} shows the number of correct fixes over four benchmarks generated by CLMs when buggy lines are given. 
For example, ``36 (-36)" shows that PLBART-base fixes 36 bugs with buggy lines provided as input, while PLBART-base fixes 72 bugs without buggy lines. 
To our surprise, all CLMs consistently fix 6\%--78\% fewer bugs when buggy lines are given. 

\begin{figure}
    \centering
    \subfigure[PLBART generates incorrect patch for NEXT\_SMALLEST in HumanEval-Java with  the buggy line given.]{
        \includegraphics[width=0.48\textwidth]{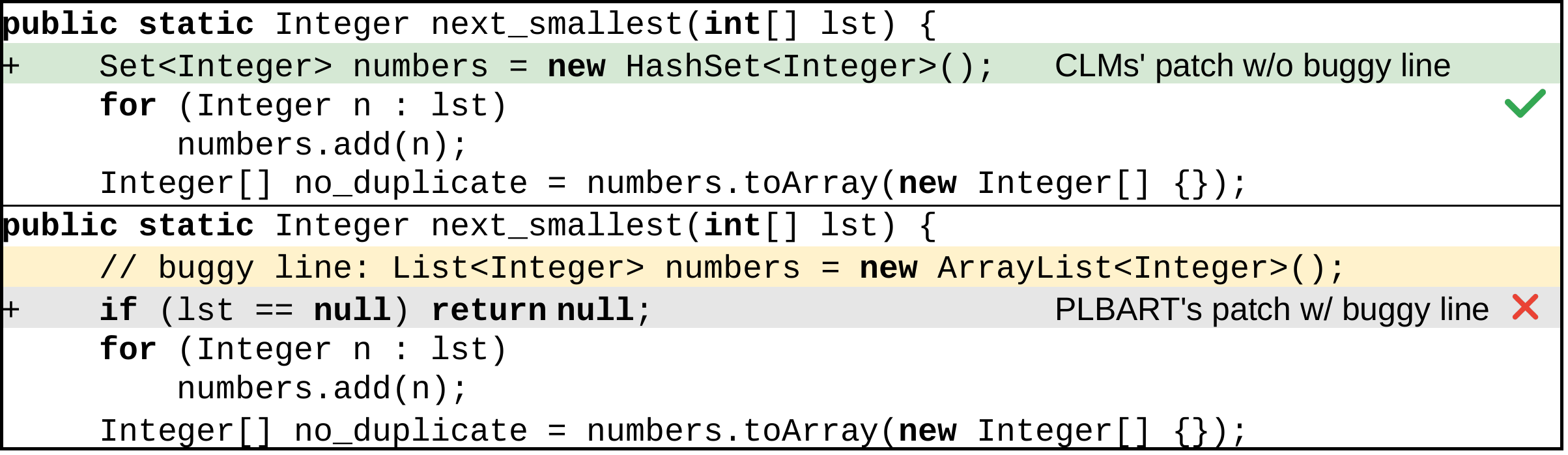}
    }
    \subfigure[CodeGen and InCoder generate incorrect patches for  FLIP\_CASE in HumanEval-Java with the buggy line given.]{
        \includegraphics[width=0.48\textwidth]{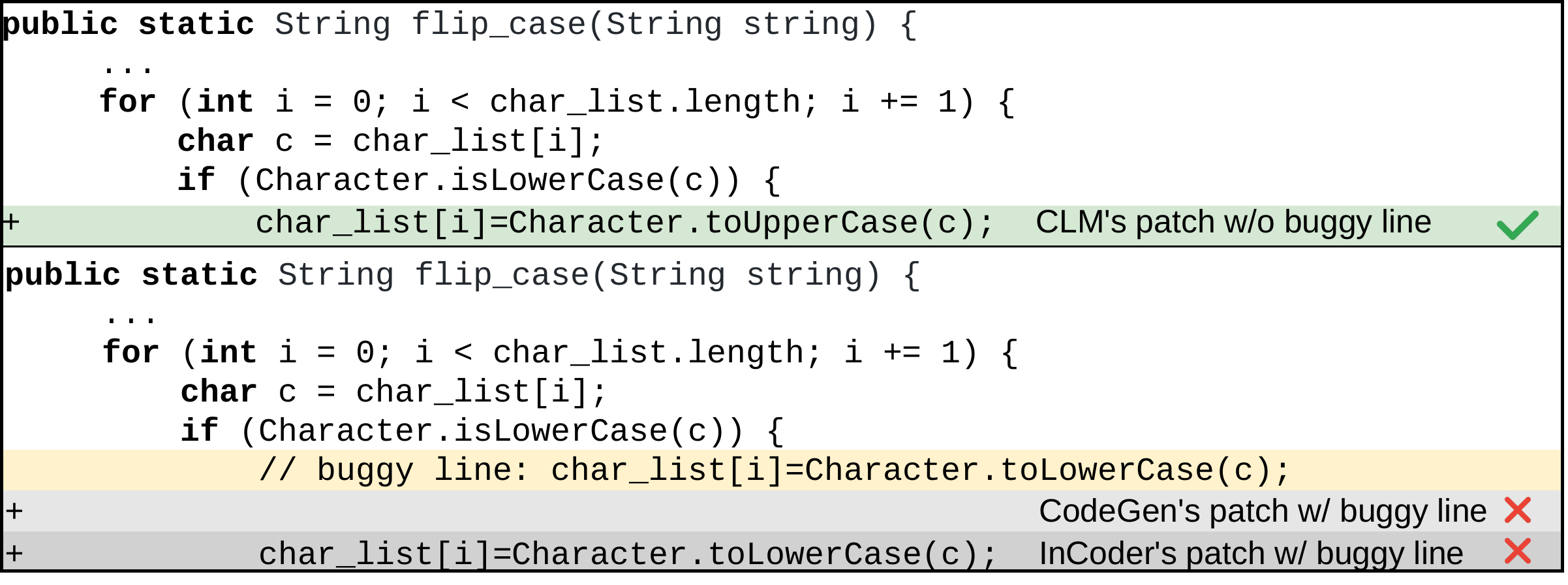}
    }
    \subfigure[CLMs fix Chart-8 in \defectsold{} only with the buggy line given.]{
        \includegraphics[width=0.48\textwidth]{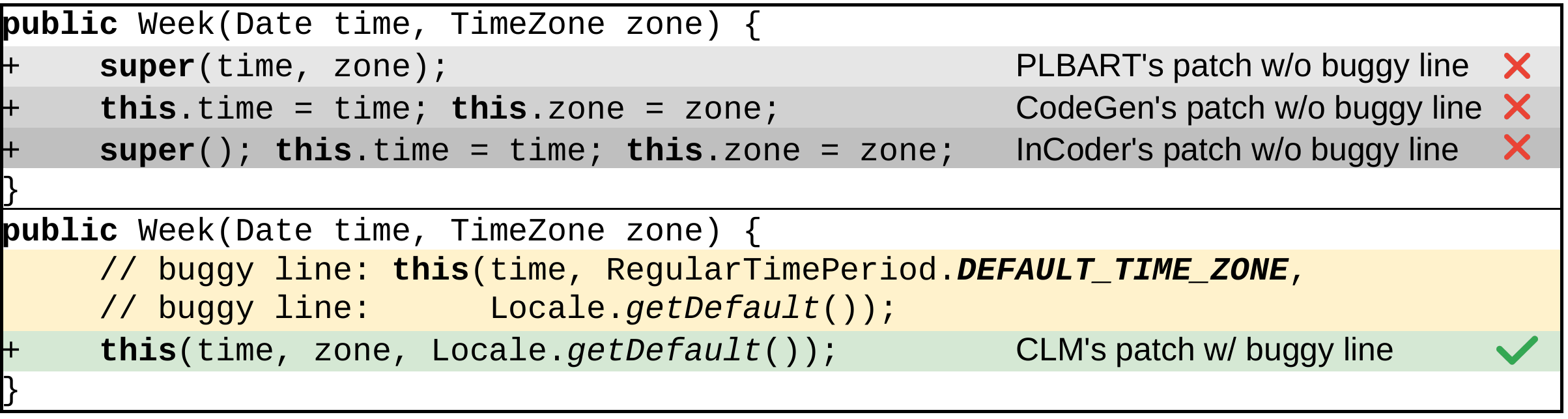}
    }
    \caption{Bug examples on which CLMs perform differently when buggy lines are given versus not given.}
    \label{fig:case lm buggyline}
\end{figure}

To understand the reason, Figure~\ref{fig:case lm buggyline}(a) shows an example, bug NEXT\_SMALLEST from HumanEval-Java, for which CLMs generate the correct patch without having the buggy line. This shows the models understand \code{no\_duplicate} in the context and thus initializes \code{numbers} as a \code{HashSet}. Yet, when the buggy line \code{List<Integer> numbers = new ArrayList<Integer>();} is given, PLBART fails to fix it anymore, generating an uncompilable patch where \code{numbers} is undeclared.
Figure~\ref{fig:case lm buggyline}(b) shows another example, FLIP\_CASE from HumanEval-Java, that CodeGen and InCoder generate incorrect patches when the buggy lines are given. CodeGen's patch deletes the whole buggy line and InCoder's patch simply repeats the buggy line, which shows that CLMs are confused by the buggy lines, and try to follow the given buggy code instead of generating the correct code. This is explainable as CLMs are not pre-trained to utilize the buggy lines.

Although CLMs fix fewer bugs when buggy lines are given, they  fix some unique bugs with the help of buggy lines. Figure~\ref{fig:case lm buggyline}(b) shows such an example, Chart-8, from \defectsold{}. Without the buggy line, CLMs' patches are incorrect for this bug as they do not have enough context to generate the correct patch. When the buggy line is given, they all can generate the correct patch \code{this(time, zone, Local.getDefault());}.

\finding{Although buggy lines enable CLMs to fix some bugs, CLMs fail to make good use of the buggy lines. CLMs generate fewer compilable patches and fix fewer bugs overall when buggy lines are given.}

\begin{table*}[ht]
    \centering
    \scriptsize
    \begin{tabular}{l|r|
    r@{\hspace{3pt}}r|
    r@{\hspace{5pt}}r@{\hspace{5pt}}r|
    r@{\hspace{5pt}}r@{\hspace{5pt}}r|
    r@{\hspace{7pt}}r|
    r@{\hspace{5pt}}r@{\hspace{5pt}}r@{\hspace{3pt}}r@{\hspace{3pt}}}
    \hline
        \textbf{Benchmarks} & \textbf{\#Bugs} &  
        \multicolumn{2}{c|}{\textbf{PLBART}} &
        \multicolumn{3}{c|}{\textbf{CodeT5}} &
        \multicolumn{3}{c|}{\textbf{CodeGen}} &
        \multicolumn{2}{c|}{\textbf{InCoder}} &
        \multicolumn{4}{c}{\textbf{DL-based APR Techniques}}\\
        & & base & large & small & base & large & 350M & 2B & 6B & 1B & 6B & CURE  & Reward & Recoder & KNOD \\
        \hline
        \hline
        \defectsold{} & 130 & 25 (12) & 29 (16) & 19 (18) & 30 (30) & 33 \textbf{(32)} & 23 (19) & 32 (21) & 38 (27) & 27 (17) & \textbf{41} (25) & 6 & 20 & 24 & 20 \\
        \defectsnew{} & 108 & 13 (4) & 17 (9) & 15 (13) & 17 (13) & 19 (18) & 20 (17) & 23 \textbf{(19)} & 23 (15) & 24 (14) & \textbf{28} (13) & 6 & 8 & 11 & 13 \\
        \quixbugs{} & 40 & 15 (4) & 17 (5) & 14 (11) & 15 (15) & 19 \textbf{(16)} & 18 (11) & 18 (3) & 18 (2) & 18 (4) & \textbf{22} (7) & 5 & 7 & 6 & 10\\
        HumanEval-Java & 164 & 41 (2) & 
        48 (-4) & 41 (38) & 54 \textbf{(49)} & 54 (48) & 52 (22) & 53 (4) & 52 (6) & 64 (24) & \textbf{70} (11) & 18 & 22 & 11 & 18\\
    \hline
    \hline
        Total & 442 & 94 (22) & 111 (26) & 89 (80) & 116 (107) & 125\textbf{(114)} & 96 (69) & 126 (47) & 131 (50) & 133 (59) & \textbf{161} (56) & 35 & 57 & 52 & 61\\
    \hline
    \end{tabular}
    \caption{Number of correct fixes generated by the ten fine-tuned CLMs and four DL-based APR techniques. Numbers in () are the improvement gained by fine-tuning. Reward stands for RewardRepair. 
    }
    \label{tab:ft lm fixing capability}
\end{table*}

\section{RQ2: Impact of Fine-tuning}

\subsection{Fixing Capabilities of Fine-tuned CLMs}
\noindent Table~\ref{tab:ft lm fixing capability} shows the number of correct fixes generated by the ten CLMs after fine-tuning. Overall, all CLMs fix more bugs after fine-tuning, with a 31\%--1,267\% improvement. As a result, fine-tuned CLMs consistently outperform DL-based APR techniques over four benchmarks. The best model, InCoder-6B, fixes 100 (164\%) more bugs than the best DL-based APR technique.


Regarding the impact of fine-tuning, CodeT5 models gain the most improvement (889\%--1,267\%) and PLBART models gain the least improvement (31\%). Although multi-task pre-training~\cite{codet5} makes CodeT5 models generate poor code before fine-tuning, they indeed learn general programming language knowledge from pre-training, which helps CodeT5 models learn great fixing capability from fine-tuning. For PLBART models, a surprising result is that the PLBART-large model fixes four fewer bugs on the HumanEval-Java benchmark after fine-tuning, which we tried to explain in Section~\ref{pre-trained vs. fine-tuned}. 

\finding{Fine-tuning with APR data improves all ten CLMs' fixing capabilities, and fine-tuned CLMs fix significantly 100 (164\%)  more bugs than the state-of-the-art DL-based APR techniques on the four benchmarks.}

\subsection{Pre-trained versus Fine-tuned CLMs}
\label{pre-trained vs. fine-tuned}
\begin{figure}
    \centering
    \subfigure[NEXT\_PALINDROME in \quixbugs{} that CLMs only fix w/ fine-tuning.]{
        \includegraphics[width=0.49\textwidth]{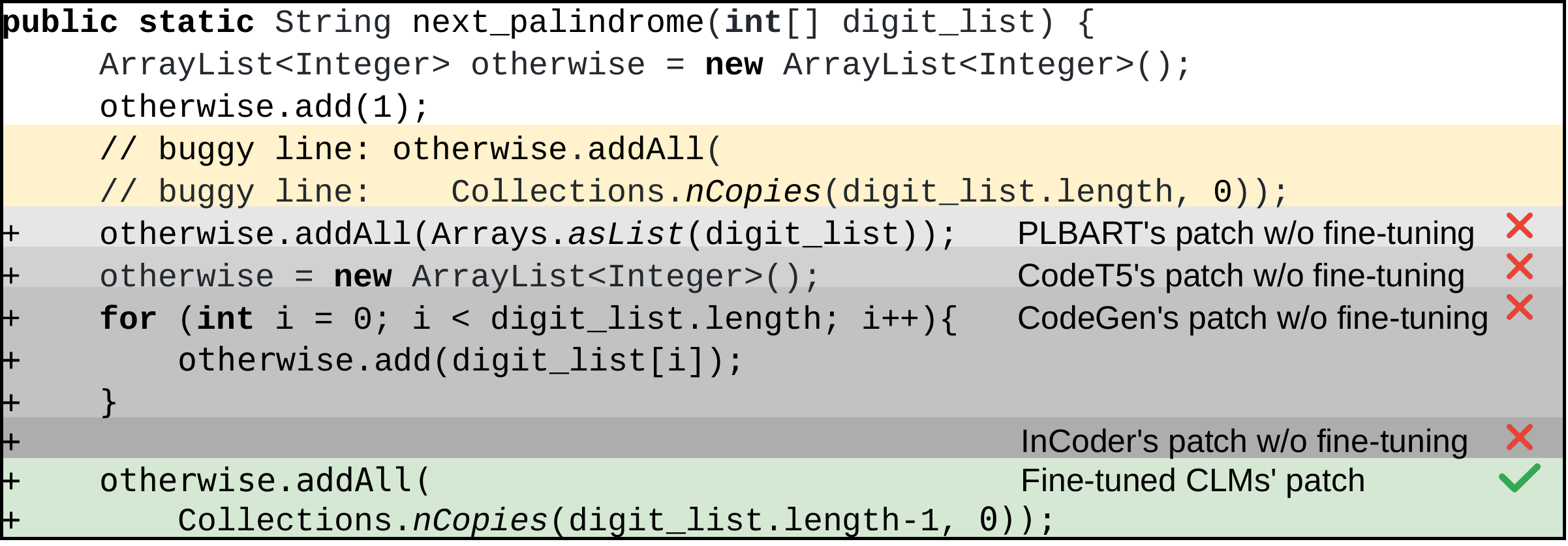}
    }
    \subfigure[SORT\_NUMBERS in HumanEval-Java that CLMs only fix w/o fine-tuning.]{
        \includegraphics[width=0.49\textwidth]{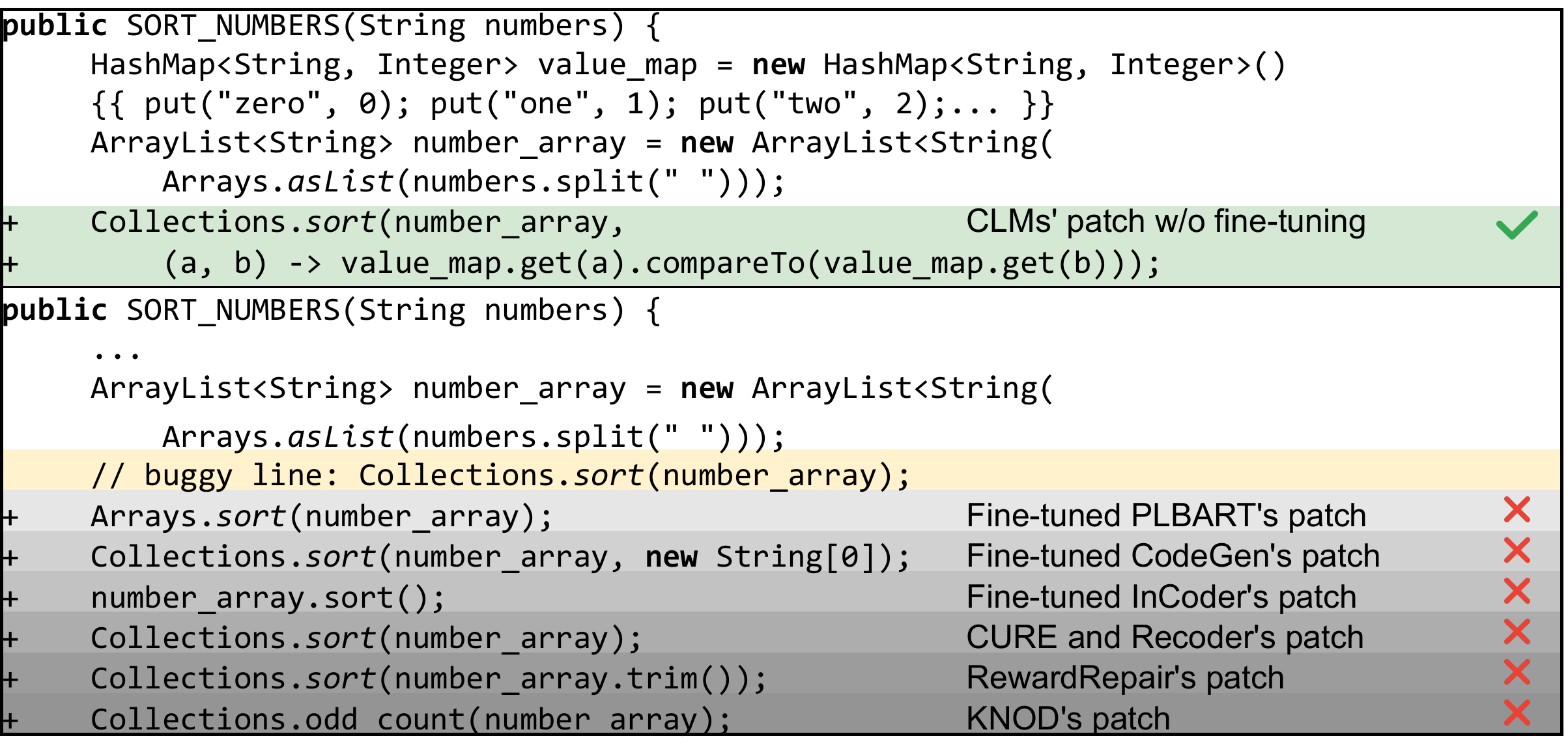}
    }
    \caption{Examples of bugs on which CLMs perform differently after fine-tuning.}
    \label{fig:case lm ft}
\end{figure}

\noindent Figure~\ref{fig:case lm ft}(a) shows an example that all the CLMs can  fix only after fine-tuning. 
Without fine-tuning, CodeT5 models generate incorrect patches that are irrelevant to the buggy line, PLBART and CodeGen models generate patches that are equivalent to the buggy lines, and InCoder models delete the buggy line. This supports our \textbf{Finding 2} that CLMs fail to utilize the buggy line information well (CLMs also fail to fix this bug without the buggy lines).
Yet, after fine-tuning, all CLMs learn to make use of the buggy lines to generate the correct patches.

Figure~\ref{fig:case lm ft} (b) shows an opposite example that CLMs can fix only without fine-tuning. PLBART, CodeGen and InCoder models fix this bug without fine-tuning and without the buggy line provided, which shows that they understand that \code{number\_array} is an array of numbers written in English, and should be sorted according to their numerical values (stored in \code{value\_map}). This reveals CLMs' strong capabilities of understanding code semantics. Yet, after fine-tuning, they all generate incorrect patches. It is surprising that the fine-tuned CLMs make a similar mistake as the DL-based APR techniques (which also fail to  fix this bug) that their patches rely on the buggy line too much, failing to figure out the target functionality from the context.

\finding{Fine-tuning with APR data enables CLMs to better leverage  buggy lines to fix more bugs. Yet, it also makes CLMs share a common shortcoming of DL-based APR techniques that they miss some bugs if they over-rely on the buggy lines. Overall, fine-tuned CLMs have the best fixing capabilities.}

\subsection{Impact of Fine-tuning Data Size}
\begin{figure}
    \centering
    \includegraphics[width=0.49\textwidth]{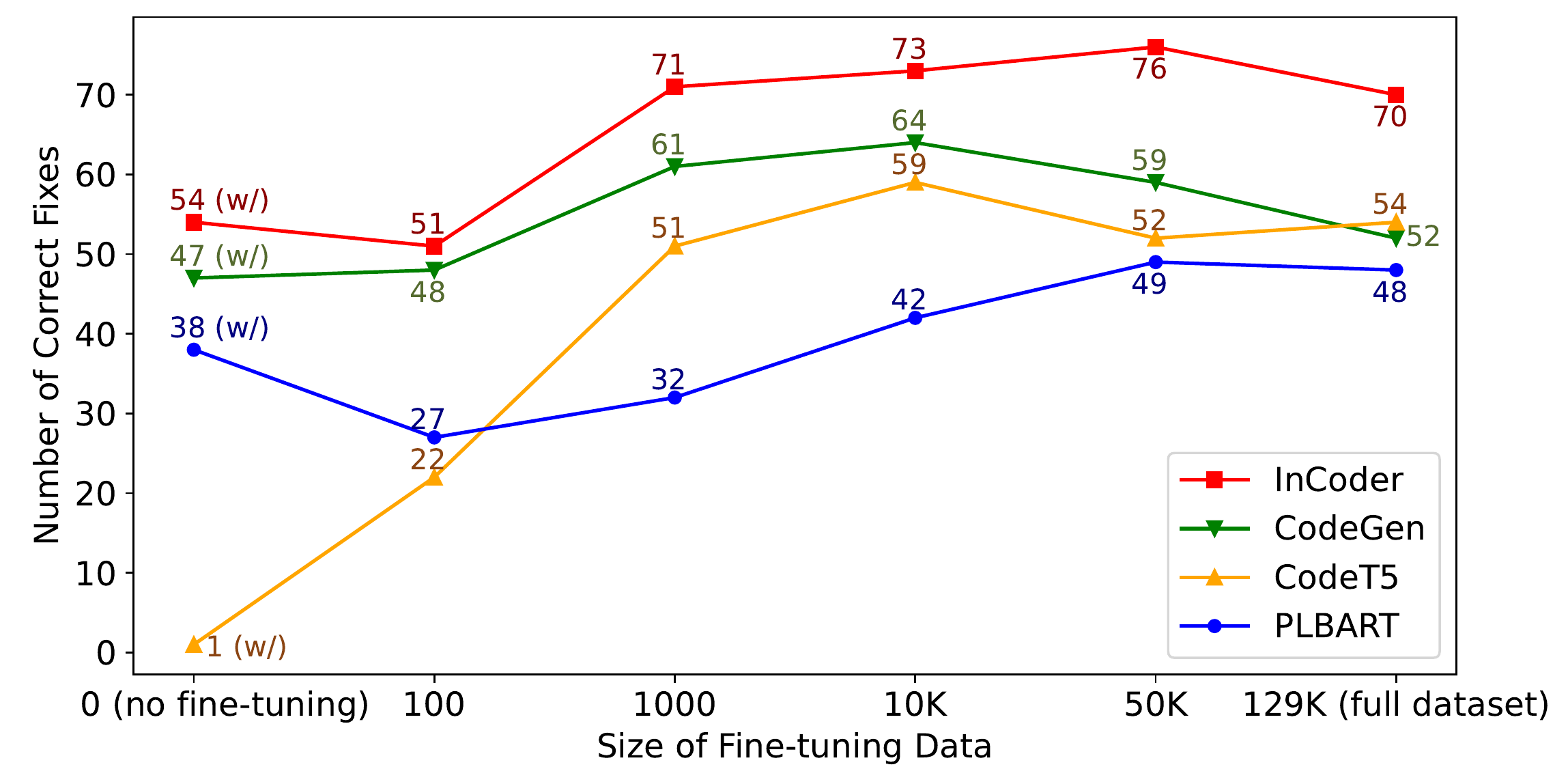}
    \caption{Number of correct fixes for HumanEval-Java generated by CLMs that are fine-tuned with different-sized APR data. (w/) stands for applying CLMs (not fine-tuned) with feeding buggy lines. 
    }
    \label{fig:fine-tuning data size}
\end{figure}

\noindent Figure~\ref{fig:fine-tuning data size} shows the number of correct fixes that CLMs generate on the HumanEval-Java benchmark when they are fine-tuned with different-sized APR data. CodeT5-large gains a great improvement after being fine-tuned with only 100 APR training instances and reaches its best fixing capability (59) after being fine-tuned with 10,000 instances. CodeGen-6B also fixes the most number of bugs after being fine-tuned with 10,000 instances. Both models share a common pattern that if the fine-tuning data increases from 10,000 to the full dataset (129,000), they start to fix fewer bugs.

InCoder-6B fixes fewer bugs after being fine-tuned with 100 APR training instances (51), but its fixing capability keeps increasing and reaches the best (76) after being fine-tuned with 50,000 instances. 

PLBART-large shows a different pattern that its fixing capability keeps a relatively stable growth as the fine-tuning data increases. Yet, the fine-tuned PLBART-large model always fixes fewer bugs than the pre-trained PLBART-large (i.e., without fine-tuning) without the buggy lines given. 

\finding{CodeT5, CodeGen, and InCoder models reach the best fixing capabilities after being fine-tuned with 10,000 and 50,000 APR instances, yet too much fine-tuning data makes them fix fewer bugs. 
The best fine-tuned PLBART model still fails to outperform the pre-trained PLBART models.}

\begin{figure*}[tb]
    \centering
    \includegraphics[width=\textwidth]{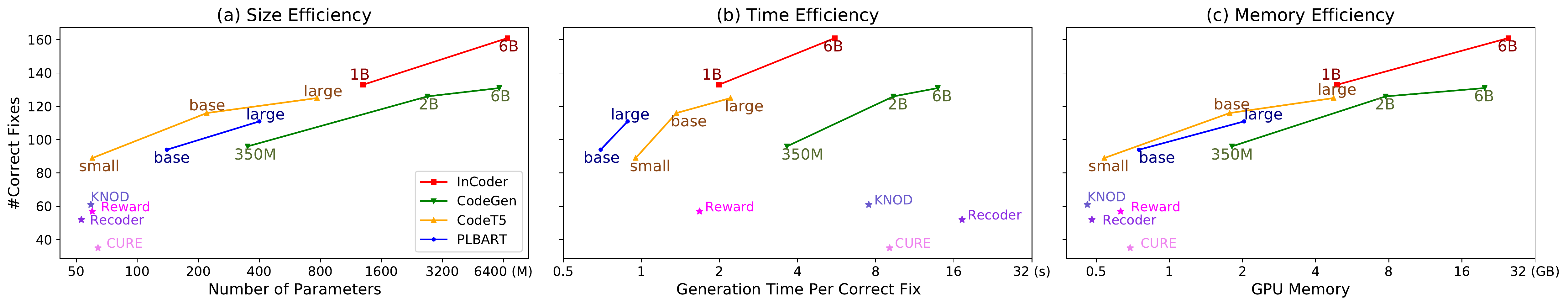}
    \caption{Size, time and memory efficiency of 
    CLMs. \lin{Is the unit ms instead of s? }}
    \label{fig:efficiency}
\end{figure*}

\section{RQ3: Size, Time, and Memory  Efficiency}
\noindent Figure~\ref{fig:efficiency} (a) shows  the  number of correct fixes that the ten fine-tuned CLMs generate as the CLMs' number of parameters grows. Larger models with more parameters consistently exhibit better fixing capability than smaller models. Fine-tuned CodeT5 and InCoder models always fix the most number of bugs compared with other models that have a similar number of parameters, i.e., CodeT5 and InCoder are the most size-efficient. Fine-tuned PLBART models have the second-best size efficiency, and CodeGen models are the least size-efficient. The result suggests that pre-training and fine-tuning larger CodeT5 or InCoder models are promising, which remains as future work since we already evaluated the largest models released.

In addition to size efficiency, we study CLMs' time efficiency. Figure~\ref{fig:efficiency} (b) shows the GPU time in seconds that each CLM takes to generate a correct fix.
PLBART models have the best time efficiency in generating patches, taking 0.70--0.89 seconds on average to generate a correct patch. CodeGen models, although fixing more bugs than PLBART models, are the least time-efficient and require 3.64--13.88 seconds on average to generate a correct patch. 

Memory requirement, or memory efficiency, is also an important consideration for practical usage. 
Figure~\ref{fig:efficiency} (c) shows the requirement of GPU memory (in GB) to apply each CLM on the four benchmarks. The CodeGen-6B and InCoder-6B models require 19.84--24.81GB GPU memory to run, which is significantly more than other CLMs. In contrast, the other CLMs can easily fit into a single card with standard 8GB or 12GB memory. Overall, CodeT5 and InCoder models always fix more bugs than PLBART and CodeGen given the same memory limitations.

We also include the size, time and memory efficiency of Dl-based APR techniques in figure~\ref{fig:efficiency}. All CLMs fix more bugs than DL-based APR techniques given the same number of parameters, time, and memory.

\finding{CodeT5 and InCoder models show the best size efficiency, thus it is more promising to develop larger CodeT5 and InCoder models than the others. PLBART, CodeT5 and InCoder models all have better time efficiency and memory efficiency than CodeGen models, and thus are better choices given limited resources. 
}

\section{Implications and Future Work}
\label{implication}

\subsection{Fine-tuning CLMs for APR}
\noindent \textbf{Improving fine-tuning:} Fine-tuned CLMs have much better fixing capabilities  than state-of-the-art DL-based APR techniques. Thus, it is promising to build future APR techniques based on CLMs instead of training from scratch. 

Yet, the fine-tuning applied in this work is straightforward and simple. Existing DL-based APR techniques have APR-specific designs, such as code-syntax guided beam search~\cite{cure}, leveraging AST structural information~\cite{recoder}, and learning execution information~\cite{rewardrepair}, which existing CLMs do not have yet. Incorporating syntax and  structural information or test cases execution information into fine-tuning may further improve the fixing capabilities of CLMs.

\smallskip \noindent \textbf{Addressing over-reliance on buggy lines:} Fine-tuned CLMs share a common shortcoming with existing DL-based APR techniques, which is the over-reliance on buggy lines. It might be caused by model biases to favor small changes to fix bugs and that fixes to most bugs in the training set are  small changes~\cite{cure}. 
\lin{is it true that most bugs only require small changes to fix? added in the training set to be safe}
Yet, this makes fixing bugs that require larger modifications to the buggy lines extra challenging. Possible solutions include balancing different bugs in the fine-tuning APR dataset, or developing separate models especially for bugs requiring big modifications. 

\subsection{Larger CodeT5 and InCoder Models}
\noindent Our Finding 6 shows that CodeT5 and InCoder models have better size efficiency. Thus, pre-training and fine-tuning larger-sized CodeT5 and InCoder models is a promising direction to fix more bugs.

\subsection{Fair and Comprehensive Evaluation}
\noindent \textbf{Improving benchmarks:} Good benchmarks are crucial for evaluating and comparing CLMs. This work releases a new benchmark HumanEval-Java that is not only more realistic than the code-refinement dataset  in CodeXGLUE for the APR task, but also not included in CLMs' pre-training data. Yet, HumanEval-Java contains mostly small programs. An APR benchmark that consists of larger buggy programs (and also not seen by CLMs) is still needed.

\smallskip \noindent \textbf{Avoiding benchmark leaking in pre-training:} CLMs rely on enormous code corpora to pre-train, which brings a threat that existing APR benchmarks such as \defects{} may be (partially) included in their pre-training datasets. It is impractical to limit the pre-training dataset of CLMs, as data is a crucial part and contribution of techniques. But we call for clearer reporting and documentation of open-source repositories used in the pre-training data of future CLM work to address the benchmark leaking problem.

\smallskip \noindent \textbf{Evaluating size, time, and memory efficiency:} In addition to the overall fixing capabilities, size efficiency shows which type of model is more promising to develop with larger sizes. Time and memory efficiency show which models perform the best given limited resources. To make more comprehensive evaluations of CLMs in future work, size, time, and memory efficiency should also be evaluated and reported.

\section{Threats to Validity and Limitations}
\noindent One threat to the evaluation of all CLM papers~\cite{plbart,codet5,codegen} is that the training data of these CLMs may contain the bugs or fixes in the four APR benchmarks since these CLMs use public repositories such as all GitHub repositories by a certain date~\cite{plbart,codet5,codegen}. This threat exists for all CLM-related papers, not just this paper. But this threat is less of a concern since CLMs do not see the pair of bugs and their fixed code during training, and their training data often contains at most the buggy code or the fixed code, but not both. We mitigate this threat by using an evaluation benchmark HumanEval-Java that has not been seen by any of the CLMs during training.\thibaud{Do we want to add that the limitation is that the moment we make HumanEval available on GitHub, this benchmark is no more useful for future mitigation of this threat.?}

We exclude the state-of-the-art code language model Codex~\cite{codex} from the results section because Codex is a black box, and one cannot fine-tune it. 
Fine-tuning Codex remains future work if Codex releases fine-tuning APIs in the future. 
We did apply Codex without fine-tuning on the four APR benchmarks. Codex correctly fixes 41 bugs from \defectsold{}, 27 bugs from \defectsnew{}, 34 bugs from \quixbugs{}, and 71 bugs from HumanEval-Java (173 in total). While Codex seems highly effective, it is particularly susceptible to the data-leaking threat, because Codex models keep updating with the latest open-source data and potential user input\footnote{https://help.openai.com/en/articles/5722486-how-your-data-is-used-to-improve-model-performance}. Making a solid and fair evaluation of Codex remains an important and challenging future work.

Another threat lies in the evaluation of patch correctness. Instead of using automated metrics such as BLEU~\cite{bleu} and CodeBLEU~\cite{codexglue}, we manually check if the patches pass the test cases are semantically equivalent to the developer patches, which could be subjective. Yet, our experiments show BLEU and CodeBLUE are indeed misleading in comparing APR techniques. The CodeBLUE score of RewardRepair's patches on the HumanEval-Java benchmark is 36.76, higher than the fine-tuned CodeT5 model (33.47), but it fixes 19 fewer bugs. We suspect this is because CLMs generate patches with more diversity, and thus better automated metrics are needed.

\section{Related Work}
\subsection{Language Models on Code Refinement}
\noindent CLMs, including~\codebert{},~\graphcodebert{},~\plbart{}, and~\codetf{}~\cite{codet5}, have been studied for code refinement.
These models are fine-tuned on the code refinement dataset offered by CodeXGLUE~\cite{codexglue}, where the input to the model is an abstracted buggy function, and the models are trained to generate the patched correct function. 
However, as we discussed in Section~\ref{intro-evaluation},(1) their performance is reported by BLEU score~\cite{bleu}, which could be misleading and different from real-world benchmarks such as~\defects~\cite{defects4j}. 
As the BLEU score only measures the similarity between the generated function with the correct function, and since no test cases are executed, the reported score cannot really tell how many bugs can be correctly fixed by these code language models. And (2) they do not study the characteristics of the programs generated by CLMs, nor study the impact of feeding buggy lines, and (3) they do not compare CLMs' fixing capabilities with DL-based APR tools. Thus, our work is different by providing a comprehensive, in-depth study of the fixing capabilities of ten CLMs, with more details about the impact of buggy lines, fine-tuning data size and their size, time, and memory efficiency.

\subsection{Automated Program Repair}
\noindent Many template-~\cite{simfix-template-1,tbar-template-2}, heuristic-~\cite{arja-heuristic-1,heuristic-2,elixir-heuristic-3}, constraint-~\cite{ACS-constraint-1,nopol-constraint-2,constraint-3}, and DL-based APR techniques~\cite{sequencer,coconut,dlfix,codit} have been developed and evaluated on~\defects and~\quixbugs benchmarks. None of these papers is fully built on large pre-trained CLMs, and they all fix fewer bugs than the three DL-based APR techniques studied in this work. Other DL-based  techniques  fix compilation bugs and syntax issues~\cite{compilation-bug-1,compilation-bug-2,compilation-bug-3} instead of runtime bugs.
Thus, the novelty, conclusion, and implications of this work are not affected.
\section{Conclusion}
\noindent This paper studies the impact that CLMs bring to the APR domain. 
We apply ten CLMs with and without fine-tuning on four APR benchmarks, including a new benchmark created in this work. 
Experiments show CLMs' competitive fixing capabilities, with the best CLM fixing 72\% more bugs than the state-of-the-art DL-based APR techniques. Fine-tuning also significantly improves CLMs, enabling CLMs to fix 31\%--1,267\% more bugs and outperform the best DL-based APR technique by 46\%--164\%. Thus, this paper shows that developing APR techniques based on CLMs, bringing APR-specific designs into the fine-tuning process of CLMs, and addressing the over-reliance on buggy lines are promising future directions to explore. In addition, this work also calls for awareness of fair and comprehensive evaluation of CLMs, including avoidance of data leaking and reporting of size, time, and memory efficiency.

\section{Data Availability}
\noindent Our replication package, including (1) the new APR benchmark HumanEval-Java, (2) the generated patches for all four benchmarks by all CLMs, (3) the fine-tuned CLM models, and (4) the source code for reproduction are available at~\cite{share}.

\section*{Acknowledgment}
\rev{\noindent We thank the reviewers for their insightful comments and suggestions. This work is partially supported by a J.P. Morgan AI Faculty Research Award. Any opinions, findings, and conclusions in this paper are those of the authors only and do not
necessarily reflect the views of our sponsors.}

\bibliographystyle{IEEEtran}
\bibliography{main}

\begin{thebibliography}{10}
\providecommand{\url}[1]{#1}
\csname url@samestyle\endcsname
\providecommand{\newblock}{\relax}
\providecommand{\bibinfo}[2]{#2}
\providecommand{\BIBentrySTDinterwordspacing}{\spaceskip=0pt\relax}
\providecommand{\BIBentryALTinterwordstretchfactor}{4}
\providecommand{\BIBentryALTinterwordspacing}{\spaceskip=\fontdimen2\font plus
\BIBentryALTinterwordstretchfactor\fontdimen3\font minus
  \fontdimen4\font\relax}
\providecommand{\BIBforeignlanguage}[2]{{%
\expandafter\ifx\csname l@#1\endcsname\relax
\typeout{** WARNING: IEEEtran.bst: No hyphenation pattern has been}%
\typeout{** loaded for the language `#1'. Using the pattern for}%
\typeout{** the default language instead.}%
\else
\language=\csname l@#1\endcsname
\fi
#2}}
\providecommand{\BIBdecl}{\relax}
\BIBdecl

\bibitem{apr-review}
M.~Monperrus, ``The living review on automated program repair,'' 2020.

\bibitem{general-apr-1}
\BIBentryALTinterwordspacing
C.~L. Goues, M.~Pradel, and A.~Roychoudhury, ``Automated program repair,''
  \emph{Commun. {ACM}}, vol.~62, no.~12, pp. 56--65, 2019. [Online]. Available:
  \url{https://doi.org/10.1145/3318162}
\BIBentrySTDinterwordspacing

\bibitem{general-apr-2}
\BIBentryALTinterwordspacing
M.~Monperrus, ``Automatic software repair: {A} bibliography,'' \emph{{ACM}
  Comput. Surv.}, vol.~51, no.~1, pp. 17:1--17:24, 2018. [Online]. Available:
  \url{https://doi.org/10.1145/3105906}
\BIBentrySTDinterwordspacing

\bibitem{sequencer}
Z.~Chen, S.~Kommrusch, M.~Tufano, L.-N. Pouchet, D.~Poshyvanyk, and
  M.~Monperrus, ``{SequenceR: Sequence-to-Sequence Learning for End-to-End
  Program Repair},'' \emph{TSE}, 2019.

\bibitem{dlfix}
Y.~Li, S.~Wang, and T.~N. Nguyen, ``{DLFix: Context-Based Code Transformation
  Learning for Automated Program Repair},'' in \emph{ICSE}.\hskip 1em plus
  0.5em minus 0.4em\relax ACM, 2020, p. 602–614.

\bibitem{coconut}
T.~Lutellier, H.~V. Pham, L.~Pang, Y.~Li, M.~Wei, and L.~Tan, ``{CoCoNuT:
  Combining Context-Aware Neural Translation Models Using Ensemble for Program
  Repair},'' in \emph{ISSTA}.\hskip 1em plus 0.5em minus 0.4em\relax ACM, 2020,
  p. 101–114.

\bibitem{cure}
N.~Jiang, T.~Lutellier, and L.~Tan, ``Cure: Code-aware neural machine
  translation for automatic program repair,'' in \emph{2021 IEEE/ACM 43rd
  International Conference on Software Engineering (ICSE)}, 2021, pp.
  1161--1173.

\bibitem{codit}
S.~Chakraborty, Y.~Ding, M.~Allamanis, and B.~Ray, ``Codit: Code editing with
  tree-based neural models,'' \emph{IEEE Transactions on Software Engineering},
  vol.~48, no.~4, pp. 1385--1399, 2022.

\bibitem{rewardrepair}
\BIBentryALTinterwordspacing
H.~Ye, M.~Martinez, and M.~Monperrus, ``Neural program repair with
  execution-based backpropagation,'' in \emph{Proceedings of the International
  Conference on Software Engineering}, 2022. [Online]. Available:
  \url{http://arxiv.org/pdf/2105.04123}
\BIBentrySTDinterwordspacing

\bibitem{recoder}
\BIBentryALTinterwordspacing
Q.~Zhu, Z.~Sun, Y.-a. Xiao, W.~Zhang, K.~Yuan, Y.~Xiong, and L.~Zhang, ``A
  syntax-guided edit decoder for neural program repair,'' in \emph{Proceedings
  of the 29th ACM Joint Meeting on European Software Engineering Conference and
  Symposium on the Foundations of Software Engineering}.\hskip 1em plus 0.5em
  minus 0.4em\relax New York, NY, USA: Association for Computing Machinery,
  2021, p. 341–353. [Online]. Available:
  \url{https://doi.org/10.1145/3468264.3468544}
\BIBentrySTDinterwordspacing

\bibitem{knod}
N.~Jiang, T.~Lutellier, Y.~Lou, L.~Tan, D.~Goldwasser, and X.~Zhang, ``Knod:
  Domain knowledge distilled tree decoder for automated program repair,'' in
  \emph{Proceedings of the International Conference on Software Engineering},
  2023.

\bibitem{transfer-vul}
\BIBentryALTinterwordspacing
Z.~Chen, S.~Kommrusch, and M.~Monperrus, ``Neural transfer learning for
  repairing security vulnerabilities in c code,'' \emph{IEEE Transactions on
  Software Engineering}, 2022. [Online]. Available:
  \url{http://arxiv.org/pdf/2104.08308}
\BIBentrySTDinterwordspacing

\bibitem{selfapr}
\BIBentryALTinterwordspacing
H.~Ye, M.~Martinez, X.~Luo, T.~Zhang, and M.~Monperrus, ``Selfapr:
  Self-supervised program repair with test execution diagnostics,'' in
  \emph{Proceedings of ASE}, 2022. [Online]. Available:
  \url{http://arxiv.org/pdf/2203.12755}
\BIBentrySTDinterwordspacing

\bibitem{simfix-template-1}
\BIBentryALTinterwordspacing
J.~Jiang, Y.~Xiong, H.~Zhang, Q.~Gao, and X.~Chen, ``Shaping program repair
  space with existing patches and similar code,'' in \emph{Proceedings of the
  27th ACM SIGSOFT International Symposium on Software Testing and Analysis},
  ser. ISSTA 2018.\hskip 1em plus 0.5em minus 0.4em\relax New York, NY, USA:
  Association for Computing Machinery, 2018, p. 298–309. [Online]. Available:
  \url{https://doi.org/10.1145/3213846.3213871}
\BIBentrySTDinterwordspacing

\bibitem{tbar-template-2}
K.~Liu, A.~Koyuncu, D.~Kim, and T.~F. Bissyand{\'e}, ``{TBar: Revisiting
  Template-Based Automated Program Repair},'' in \emph{ISSTA}.\hskip 1em plus
  0.5em minus 0.4em\relax ACM, 2019.

\bibitem{arja-heuristic-1}
\BIBentryALTinterwordspacing
Y.~Yuan and W.~Banzhaf, ``{ARJA:} automated repair of java programs via
  multi-objective genetic programming,'' \emph{{IEEE} Trans. Software Eng.},
  vol.~46, no.~10, pp. 1040--1067, 2020. [Online]. Available:
  \url{https://doi.org/10.1109/TSE.2018.2874648}
\BIBentrySTDinterwordspacing

\bibitem{heuristic-2}
\BIBentryALTinterwordspacing
M.~Wen, J.~Chen, R.~Wu, D.~Hao, and S.~Cheung, ``Context-aware patch generation
  for better automated program repair,'' in \emph{Proceedings of the 40th
  International Conference on Software Engineering, {ICSE} 2018, Gothenburg,
  Sweden, May 27 - June 03, 2018}, M.~Chaudron, I.~Crnkovic, M.~Chechik, and
  M.~Harman, Eds.\hskip 1em plus 0.5em minus 0.4em\relax {ACM}, 2018, pp.
  1--11. [Online]. Available: \url{https://doi.org/10.1145/3180155.3180233}
\BIBentrySTDinterwordspacing

\bibitem{elixir-heuristic-3}
\BIBentryALTinterwordspacing
R.~K. Saha, Y.~Lyu, H.~Yoshida, and M.~R. Prasad, ``{ELIXIR:} effective object
  oriented program repair,'' in \emph{Proceedings of the 32nd {IEEE/ACM}
  International Conference on Automated Software Engineering, {ASE} 2017,
  Urbana, IL, USA, October 30 - November 03, 2017}, G.~Rosu, M.~D. Penta, and
  T.~N. Nguyen, Eds.\hskip 1em plus 0.5em minus 0.4em\relax {IEEE} Computer
  Society, 2017, pp. 648--659. [Online]. Available:
  \url{https://doi.org/10.1109/ASE.2017.8115675}
\BIBentrySTDinterwordspacing

\bibitem{ACS-constraint-1}
Y.~Xiong, J.~Wang, R.~Yan, J.~Zhang, S.~Han, G.~Huang, and L.~Zhang, ``Precise
  condition synthesis for program repair,'' in \emph{2017 IEEE/ACM 39th
  International Conference on Software Engineering (ICSE)}, 2017, pp. 416--426.

\bibitem{nopol-constraint-2}
\BIBentryALTinterwordspacing
J.~Xuan, M.~Martinez, F.~Demarco, M.~Clement, S.~R.~L. Marcote, T.~Durieux,
  D.~L. Berre, and M.~Monperrus, ``Nopol: Automatic repair of conditional
  statement bugs in java programs,'' \emph{{IEEE} Trans. Software Eng.},
  vol.~43, no.~1, pp. 34--55, 2017. [Online]. Available:
  \url{https://doi.org/10.1109/TSE.2016.2560811}
\BIBentrySTDinterwordspacing

\bibitem{constraint-3}
\BIBentryALTinterwordspacing
M.~Martinez and M.~Monperrus, ``Ultra-large repair search space with
  automatically mined templates: The cardumen mode of astor,'' in
  \emph{Search-Based Software Engineering - 10th International Symposium,
  {SSBSE} 2018, Montpellier, France, September 8-9, 2018, Proceedings}, ser.
  Lecture Notes in Computer Science, T.~E. Colanzi and P.~McMinn, Eds., vol.
  11036.\hskip 1em plus 0.5em minus 0.4em\relax Springer, 2018, pp. 65--86.
  [Online]. Available: \url{https://doi.org/10.1007/978-3-319-99241-9\_3}
\BIBentrySTDinterwordspacing

\bibitem{trust-apr}
\BIBentryALTinterwordspacing
Y.~Noller, R.~Shariffdeen, X.~Gao, and A.~Roychoudhury, ``Trust enhancement
  issues in program repair,'' in \emph{Proceedings of the 44th International
  Conference on Software Engineering}, ser. ICSE '22.\hskip 1em plus 0.5em
  minus 0.4em\relax New York, NY, USA: Association for Computing Machinery,
  2022, p. 2228–2240. [Online]. Available:
  \url{https://doi.org/10.1145/3510003.3510040}
\BIBentrySTDinterwordspacing

\bibitem{plbart}
\BIBentryALTinterwordspacing
W.~Ahmad, S.~Chakraborty, B.~Ray, and K.-W. Chang, ``Unified pre-training for
  program understanding and generation,'' in \emph{Proceedings of the 2021
  Conference of the North American Chapter of the Association for Computational
  Linguistics: Human Language Technologies}.\hskip 1em plus 0.5em minus
  0.4em\relax Online: Association for Computational Linguistics, Jun. 2021, pp.
  2655--2668. [Online]. Available:
  \url{https://aclanthology.org/2021.naacl-main.211}
\BIBentrySTDinterwordspacing

\bibitem{codet5}
W.~Yue, W.~Weishi, J.~Shafiq, and C.~H. Steven, ``Codet5: Identifier-aware
  unified pre-trained encoder-decoder models for code understanding and
  generation,'' in \emph{Proceedings of the 2021 Conference on Empirical
  Methods in Natural Language Processing, EMNLP 2021}, 2021.

\bibitem{gpt-1}
A.~Radford, K.~Narasimhan, T.~Salimans, and I.~Sutskever, ``Improving language
  understanding by generative pre-training,'' 2018.

\bibitem{gpt-2}
A.~Radford, J.~Wu, R.~Child, D.~Luan, D.~Amodei, and I.~Sutskever, ``Language
  models are unsupervised multitask learners,'' 2019.

\bibitem{gpt-3}
\BIBentryALTinterwordspacing
T.~B. Brown, B.~Mann, N.~Ryder, M.~Subbiah, J.~Kaplan, P.~Dhariwal,
  A.~Neelakantan, P.~Shyam, G.~Sastry, A.~Askell, S.~Agarwal,
  A.~Herbert{-}Voss, G.~Krueger, T.~Henighan, R.~Child, A.~Ramesh, D.~M.
  Ziegler, J.~Wu, C.~Winter, C.~Hesse, M.~Chen, E.~Sigler, M.~Litwin, S.~Gray,
  B.~Chess, J.~Clark, C.~Berner, S.~McCandlish, A.~Radford, I.~Sutskever, and
  D.~Amodei, ``Language models are few-shot learners,'' \emph{CoRR}, vol.
  abs/2005.14165, 2020. [Online]. Available:
  \url{https://arxiv.org/abs/2005.14165}
\BIBentrySTDinterwordspacing

\bibitem{gpt-j}
B.~Wang and A.~Komatsuzaki, ``{GPT-J-6B: A 6 Billion Parameter Autoregressive
  Language Model},'' May 2021.

\bibitem{gpt-neo}
\BIBentryALTinterwordspacing
S.~Black, G.~Leo, P.~Wang, C.~Leahy, and S.~Biderman, ``{GPT-Neo: Large Scale
  Autoregressive Language Modeling with Mesh-Tensorflow},'' Mar. 2021.
  [Online]. Available: \url{https://doi.org/10.5281/zenodo.5297715}
\BIBentrySTDinterwordspacing

\bibitem{bert}
\BIBentryALTinterwordspacing
J.~Devlin, M.~Chang, K.~Lee, and K.~Toutanova, ``{BERT:} pre-training of deep
  bidirectional transformers for language understanding,'' \emph{CoRR}, vol.
  abs/1810.04805, 2018. [Online]. Available:
  \url{http://arxiv.org/abs/1810.04805}
\BIBentrySTDinterwordspacing

\bibitem{roberta}
\BIBentryALTinterwordspacing
Y.~Liu, M.~Ott, N.~Goyal, J.~Du, M.~Joshi, D.~Chen, O.~Levy, M.~Lewis,
  L.~Zettlemoyer, and V.~Stoyanov, ``Roberta: {A} robustly optimized {BERT}
  pretraining approach,'' \emph{CoRR}, vol. abs/1907.11692, 2019. [Online].
  Available: \url{http://arxiv.org/abs/1907.11692}
\BIBentrySTDinterwordspacing

\bibitem{bart}
\BIBentryALTinterwordspacing
M.~Lewis, Y.~Liu, N.~Goyal, M.~Ghazvininejad, A.~Mohamed, O.~Levy, V.~Stoyanov,
  and L.~Zettlemoyer, ``{BART:} denoising sequence-to-sequence pre-training for
  natural language generation, translation, and comprehension,'' \emph{CoRR},
  vol. abs/1910.13461, 2019. [Online]. Available:
  \url{http://arxiv.org/abs/1910.13461}
\BIBentrySTDinterwordspacing

\bibitem{graph-bert}
\BIBentryALTinterwordspacing
J.~Zhang, H.~Zhang, C.~Xia, and L.~Sun, ``Graph-bert: Only attention is needed
  for learning graph representations,'' \emph{CoRR}, vol. abs/2001.05140, 2020.
  [Online]. Available: \url{https://arxiv.org/abs/2001.05140}
\BIBentrySTDinterwordspacing

\bibitem{turing-nlg}
\BIBentryALTinterwordspacing
S.~Smith, M.~Patwary, B.~Norick, P.~LeGresley, S.~Rajbhandari, J.~Casper,
  Z.~Liu, S.~Prabhumoye, G.~Zerveas, V.~Korthikanti, E.~Zheng, R.~Child, R.~Y.
  Aminabadi, J.~Bernauer, X.~Song, M.~Shoeybi, Y.~He, M.~Houston, S.~Tiwary,
  and B.~Catanzaro, ``Using deepspeed and megatron to train megatron-turing
  {NLG} 530b, {A} large-scale generative language model,'' \emph{CoRR}, vol.
  abs/2201.11990, 2022. [Online]. Available:
  \url{https://arxiv.org/abs/2201.11990}
\BIBentrySTDinterwordspacing

\bibitem{codegen}
E.~Nijkamp, B.~Pang, H.~Hayashi, L.~Tu, H.~Wang, Y.~Zhou, S.~Savarese, and
  C.~Xiong, ``A conversational paradigm for program synthesis,'' \emph{arXiv
  preprint}, 2022.

\bibitem{codex}
\BIBentryALTinterwordspacing
M.~Chen, J.~Tworek, H.~Jun, Q.~Yuan, H.~P. de~Oliveira~Pinto, J.~Kaplan,
  H.~Edwards, Y.~Burda, N.~Joseph, G.~Brockman, A.~Ray, R.~Puri, G.~Krueger,
  M.~Petrov, H.~Khlaaf, G.~Sastry, P.~Mishkin, B.~Chan, S.~Gray, N.~Ryder,
  M.~Pavlov, A.~Power, L.~Kaiser, M.~Bavarian, C.~Winter, P.~Tillet, F.~P.
  Such, D.~Cummings, M.~Plappert, F.~Chantzis, E.~Barnes, A.~Herbert{-}Voss,
  W.~H. Guss, A.~Nichol, A.~Paino, N.~Tezak, J.~Tang, I.~Babuschkin, S.~Balaji,
  S.~Jain, W.~Saunders, C.~Hesse, A.~N. Carr, J.~Leike, J.~Achiam, V.~Misra,
  E.~Morikawa, A.~Radford, M.~Knight, M.~Brundage, M.~Murati, K.~Mayer,
  P.~Welinder, B.~McGrew, D.~Amodei, S.~McCandlish, I.~Sutskever, and
  W.~Zaremba, ``Evaluating large language models trained on code,''
  \emph{CoRR}, vol. abs/2107.03374, 2021. [Online]. Available:
  \url{https://arxiv.org/abs/2107.03374}
\BIBentrySTDinterwordspacing

\bibitem{codebert}
\BIBentryALTinterwordspacing
Z.~Feng, D.~Guo, D.~Tang, N.~Duan, X.~Feng, M.~Gong, L.~Shou, B.~Qin, T.~Liu,
  D.~Jiang, and M.~Zhou, ``Codebert: {A} pre-trained model for programming and
  natural languages,'' \emph{CoRR}, vol. abs/2002.08155, 2020. [Online].
  Available: \url{https://arxiv.org/abs/2002.08155}
\BIBentrySTDinterwordspacing

\bibitem{graph-codebert}
\BIBentryALTinterwordspacing
D.~Guo, S.~Ren, S.~Lu, Z.~Feng, D.~Tang, S.~Liu, L.~Zhou, N.~Duan,
  A.~Svyatkovskiy, S.~Fu, M.~Tufano, S.~K. Deng, C.~B. Clement, D.~Drain,
  N.~Sundaresan, J.~Yin, D.~Jiang, and M.~Zhou, ``Graphcodebert: Pre-training
  code representations with data flow,'' \emph{CoRR}, vol. abs/2009.08366,
  2020. [Online]. Available: \url{https://arxiv.org/abs/2009.08366}
\BIBentrySTDinterwordspacing

\bibitem{codexglue}
S.~Lu, D.~Guo, S.~Ren, J.~Huang, A.~Svyatkovskiy, A.~Blanco, C.~B. Clement,
  D.~Drain, D.~Jiang, D.~Tang, G.~Li, L.~Zhou, L.~Shou, L.~Zhou, M.~Tufano,
  M.~Gong, M.~Zhou, N.~Duan, N.~Sundaresan, S.~K. Deng, S.~Fu, and S.~Liu,
  ``Codexglue: {A} machine learning benchmark dataset for code understanding
  and generation,'' \emph{CoRR}, vol. abs/2102.04664, 2021.

\bibitem{defects4j}
R.~Just, D.~Jalali, and M.~D. Ernst, ``{Defects4J: A Database of Existing
  Faults to Enable Controlled Testing Studies for Java Programs},'' in
  \emph{ISSTA}, 2014, pp. 437--440.

\bibitem{quixbugs}
D.~Lin, J.~Koppel, A.~Chen, and A.~Solar-Lezama, ``{QuixBugs: A Multi-Lingual
  Program Repair Benchmark Set Based on the Quixey Challenge},'' in
  \emph{SPLASH}, 2017, p. 55–56.

\bibitem{codex-quixbugs}
J.~A. Prenner, H.~Babii, and R.~Robbes, ``Can openai's codex fix bugs?: An
  evaluation on quixbugs,'' in \emph{2022 IEEE/ACM International Workshop on
  Automated Program Repair (APR)}, 2022, pp. 69--75.

\bibitem{barz2020we}
B.~Barz and J.~Denzler, ``Do we train on test data? purging cifar of
  near-duplicates,'' \emph{Journal of Imaging}, vol.~6, no.~6, p.~41, 2020.

\bibitem{tan2015online}
M.~Tan, L.~Tan, S.~Dara, and C.~Mayeux, ``Online defect prediction for
  imbalanced data,'' in \emph{2015 IEEE/ACM 37th IEEE International Conference
  on Software Engineering}, vol.~2.\hskip 1em plus 0.5em minus 0.4em\relax
  IEEE, 2015, pp. 99--108.

\bibitem{incoder}
\BIBentryALTinterwordspacing
D.~Fried, A.~Aghajanyan, J.~Lin, S.~Wang, E.~Wallace, F.~Shi, R.~Zhong, W.-t.
  Yih, L.~Zettlemoyer, and M.~Lewis, ``Incoder: A generative model for code
  infilling and synthesis,'' 2022. [Online]. Available:
  \url{https://arxiv.org/abs/2204.05999}
\BIBentrySTDinterwordspacing

\bibitem{T5}
\BIBentryALTinterwordspacing
C.~Raffel, N.~Shazeer, A.~Roberts, K.~Lee, S.~Narang, M.~Matena, Y.~Zhou,
  W.~Li, and P.~J. Liu, ``Exploring the limits of transfer learning with a
  unified text-to-text transformer,'' \emph{CoRR}, vol. abs/1910.10683, 2019.
  [Online]. Available: \url{http://arxiv.org/abs/1910.10683}
\BIBentrySTDinterwordspacing

\bibitem{data-size-1}
\BIBentryALTinterwordspacing
E.-S.~A. Lee, S.~Thillainathan, S.~Nayak, S.~Ranathunga, D.~I. Adelani, R.~Su,
  and A.~D. McCarthy, ``Pre-trained multilingual sequence-to-sequence models: A
  hope for low-resource language translation?'' 2022. [Online]. Available:
  \url{https://arxiv.org/abs/2203.08850}
\BIBentrySTDinterwordspacing

\bibitem{data-size-2}
\BIBentryALTinterwordspacing
H.~Mehrafarin, S.~Rajaee, and M.~T. Pilehvar, ``On the importance of data size
  in probing fine-tuned models,'' in \emph{Findings of the Association for
  Computational Linguistics: ACL 2022}.\hskip 1em plus 0.5em minus 0.4em\relax
  Dublin, Ireland: Association for Computational Linguistics, May 2022, pp.
  228--238. [Online]. Available:
  \url{https://aclanthology.org/2022.findings-acl.20}
\BIBentrySTDinterwordspacing

\bibitem{transformer}
\BIBentryALTinterwordspacing
A.~Vaswani, N.~Shazeer, N.~Parmar, J.~Uszkoreit, L.~Jones, A.~N. Gomez,
  L.~Kaiser, and I.~Polosukhin, ``Attention is all you need,'' \emph{CoRR},
  vol. abs/1706.03762, 2017. [Online]. Available:
  \url{http://arxiv.org/abs/1706.03762}
\BIBentrySTDinterwordspacing

\bibitem{codesearchnet}
\BIBentryALTinterwordspacing
H.~Husain, H.~Wu, T.~Gazit, M.~Allamanis, and M.~Brockschmidt, ``Codesearchnet
  challenge: Evaluating the state of semantic code search,'' \emph{CoRR}, vol.
  abs/1909.09436, 2019. [Online]. Available:
  \url{http://arxiv.org/abs/1909.09436}
\BIBentrySTDinterwordspacing

\bibitem{thepile}
\BIBentryALTinterwordspacing
L.~Gao, S.~Biderman, S.~Black, L.~Golding, T.~Hoppe, C.~Foster, J.~Phang,
  H.~He, A.~Thite, N.~Nabeshima, S.~Presser, and C.~Leahy, ``The pile: An 800gb
  dataset of diverse text for language modeling,'' \emph{CoRR}, vol.
  abs/2101.00027, 2021. [Online]. Available:
  \url{https://arxiv.org/abs/2101.00027}
\BIBentrySTDinterwordspacing

\bibitem{xglm}
\BIBentryALTinterwordspacing
X.~V. Lin, T.~Mihaylov, M.~Artetxe, T.~Wang, S.~Chen, D.~Simig, M.~Ott,
  N.~Goyal, S.~Bhosale, J.~Du, R.~Pasunuru, S.~Shleifer, P.~S. Koura,
  V.~Chaudhary, B.~O'Horo, J.~Wang, L.~Zettlemoyer, Z.~Kozareva, M.~T. Diab,
  V.~Stoyanov, and X.~Li, ``Few-shot learning with multilingual language
  models,'' \emph{CoRR}, vol. abs/2112.10668, 2021. [Online]. Available:
  \url{https://arxiv.org/abs/2112.10668}
\BIBentrySTDinterwordspacing

\bibitem{huggingface}
\BIBentryALTinterwordspacing
T.~Wolf, L.~Debut, V.~Sanh, J.~Chaumond, C.~Delangue, A.~Moi, P.~Cistac,
  T.~Rault, R.~Louf, M.~Funtowicz, J.~Davison, S.~Shleifer, P.~von Platen,
  C.~Ma, Y.~Jernite, J.~Plu, C.~Xu, T.~L. Scao, S.~Gugger, M.~Drame, Q.~Lhoest,
  and A.~M. Rush, ``Transformers: State-of-the-art natural language
  processing,'' in \emph{Proceedings of the 2020 Conference on Empirical
  Methods in Natural Language Processing: System Demonstrations}.\hskip 1em
  plus 0.5em minus 0.4em\relax Online: Association for Computational
  Linguistics, Oct. 2020, pp. 38--45. [Online]. Available:
  \url{https://www.aclweb.org/anthology/2020.emnlp-demos.6}
\BIBentrySTDinterwordspacing

\bibitem{alpharepair}
\BIBentryALTinterwordspacing
C.~S. Xia and L.~Zhang, ``Less training, more repairing please: Revisiting
  automated program repair via zero-shot learning,'' in \emph{Proceedings of
  the 30th ACM Joint European Software Engineering Conference and Symposium on
  the Foundations of Software Engineering}, ser. ESEC/FSE 2022.\hskip 1em plus
  0.5em minus 0.4em\relax New York, NY, USA: Association for Computing
  Machinery, 2022, p. 959–971. [Online]. Available:
  \url{https://doi.org/10.1145/3540250.3549101}
\BIBentrySTDinterwordspacing

\bibitem{fairseq}
M.~Ott, S.~Edunov, A.~Baevski, A.~Fan, S.~Gross, N.~Ng, D.~Grangier, and
  M.~Auli, ``fairseq: A fast, extensible toolkit for sequence modeling,'' in
  \emph{Proceedings of NAACL-HLT 2019: Demonstrations}, 2019.

\bibitem{graph-transfrmer-1}
\BIBentryALTinterwordspacing
V.~P. Dwivedi and X.~Bresson, ``A generalization of transformer networks to
  graphs,'' \emph{CoRR}, vol. abs/2012.09699, 2020. [Online]. Available:
  \url{https://arxiv.org/abs/2012.09699}
\BIBentrySTDinterwordspacing

\bibitem{firstlogic}
\BIBentryALTinterwordspacing
Z.~Hu, X.~Ma, Z.~Liu, E.~Hovy, and E.~Xing, ``Harnessing deep neural networks
  with logic rules,'' in \emph{Proceedings of the 54th Annual Meeting of the
  Association for Computational Linguistics (Volume 1: Long Papers)}.\hskip 1em
  plus 0.5em minus 0.4em\relax Berlin, Germany: Association for Computational
  Linguistics, Aug. 2016, pp. 2410--2420. [Online]. Available:
  \url{https://aclanthology.org/P16-1228}
\BIBentrySTDinterwordspacing

\bibitem{bleu}
\BIBentryALTinterwordspacing
C.-Y. Lin and F.~J. Och, ``{ORANGE}: a method for evaluating automatic
  evaluation metrics for machine translation,'' in \emph{{COLING} 2004:
  Proceedings of the 20th International Conference on Computational
  Linguistics}.\hskip 1em plus 0.5em minus 0.4em\relax Geneva, Switzerland:
  COLING, aug 23{--}aug 27 2004, pp. 501--507. [Online]. Available:
  \url{https://aclanthology.org/C04-1072}
\BIBentrySTDinterwordspacing

\bibitem{compilation-bug-1}
A.~Mesbah, A.~Rice, E.~Johnston, N.~Glorioso, and E.~Aftandilian, ``{DeepDelta:
  Learning to Repair Compilation Errors},'' in \emph{ESEC/FSE}.\hskip 1em plus
  0.5em minus 0.4em\relax ACM, 2019, pp. 925--936.

\bibitem{compilation-bug-2}
R.~Gupta, S.~Pal, A.~Kanade, and S.~Shevade, ``{Deepfix: Fixing Common C
  Language Errors by Deep Learning},'' in \emph{AAAI}, 2017, pp. 1345--1351.

\bibitem{compilation-bug-3}
E.~A. Santos, J.~C. Campbell, A.~Hindle, and J.~N. Amaral, ``{Finding and
  Correcting Syntax Errors Using Recurrent Neural Networks},'' \emph{PeerJ
  PrePrints}, vol.~5, p. e3123v1, 2017.

\bibitem{share}
\BIBentryALTinterwordspacing
``Replication package of this work,'' 2022. [Online]. Available:
  \url{https://github.com/lin-tan/clm}
\BIBentrySTDinterwordspacing

\end{thebibliography}

\end{document}